\documentclass[a4paper,11pt]{article}
\pdfoutput=1 

\usepackage{jcappub} 

\usepackage[T1]{fontenc} 
\usepackage{placeins}
\usepackage[utf8]{inputenc}
\usepackage{amsmath,mleftright}
\usepackage{xparse}
\usepackage{caption}
\usepackage{subcaption}
\NewDocumentCommand{\evalat}{sO{\big}mm}{%
  \IfBooleanTF{#1}
   {\mleft. #3 \mright|_{#4}}
   {#3#2|_{#4}}%
}

\usepackage{graphicx}
\usepackage{dcolumn}
\usepackage{amssymb}
\usepackage{amsmath}
\usepackage{color}
\usepackage{verbatim}
\usepackage{multirow}
\usepackage{soul}
\usepackage{parskip}
\usepackage{tabularx}

\graphicspath{
{./}
{./figure/}
{./figure/DIC/}
{./figure/Fisher_foreground_model_comparison/}
{./figure/Fisher_MCMC_comparison/}
}

\def\al{\alpha} 
\def\be{\beta}

\def\th{\theta}

\newcommand{\ben}{\begin{equation}}
\newcommand{\een}{\end{equation}}
\newcommand{\bea}{\begin{eqnarray}}
\newcommand{\eea}{\end{eqnarray}}
\newcommand{\ba}{\begin{array}}
\newcommand{\ea}{\end{array}}
\newcommand{\bit}{\begin{itemize}}
\newcommand{\eit}{\end{itemize}}

\newcommand{\X}{\mit{X}}
\newcommand{\Y}{\mit{Y}}
\newcommand{\Z}{\mit{Z}}

\newcommand{\A}{\mit{A}}
\newcommand{\E}{\mit{E}}
\newcommand{\T}{\mit{T}}
\newcommand{\alOne}{\alpha_\text{1}} 
\newcommand{\alTwo}{\alpha_\text{2}} 
\newcommand{\AOne}{A_\text{1}} 
\newcommand{\ATwo}{A_\text{2}} 
\newcommand{\alAstro}{\alpha_{\rm{ECB}}} 

\newcommand{\Aastro}{A_{\rm{ECB}}} 

\newcommand{\cs}{c_\text{s}} 

\newcommand{\fp}{f_\text{p}}

\newcommand{\Nacc}{N_{\text{acc}}} 
\newcommand{\Npos}{N_{\text{pos}}} 

\newcommand{\OmGW}{\Omega_\text{gw}} 
\newcommand{\OmGWPT}{\Omega_\text{PT}}

\newcommand{\OmPeak}{\Omega_\text{p}}

\newcommand{\Omecb}{\Omega_{\text{ECB}}} 

\newcommand{\OmDWD}{\Omega_{\text{DWD}}} 
\newcommand{\OmPT}{\Omega_{\text{PT}}} 

\newcommand{\rb}{r_\text{b}} 

\newcommand{\Tn}{T_\text{n}} 

\newcommand{\vw}{v_\text{w}} 

\definecolor{darkgreen}{rgb}{0.0,0.5,0.0}

\title{Prospects for LISA to detect a gravitational-wave background from first order phase transitions}
\subheader{HIP-2022-22/TH}
\author[a,b]{Guillaume Boileau,} 
\author[a]{Nelson Christensen,} 
\author[c,d]{Chloe Gowling,} 
\author[d,c]{Mark Hindmarsh,}
\author[e]{Renate Meyer.} 

\affiliation[a]{Artemis, Observatoire de la C\^{o}te d'Azur, Universit\'{e} C\^{o}te d'Azur, CNRS, CS 34229, F-06304 Nice Cedex 4, France}
\affiliation[b]{Universiteit Antwerpen, Prinsstraat 13, 2000 Antwerpen, Belgium}

\affiliation[c]{Department of Physics and Astronomy, University of Sussex, BN1 9QH, Brighton, UK}
\affiliation[d]{Department of Physics and Helsinki Institute of Physics, PL 64, FI-00014 University of Helsinki, Finland}

\affiliation[e]{ Department of Statistics, University of Auckland, Auckland, New Zealand} 

\emailAdd{guillaume.boileau@oca.eu}
\emailAdd{guillaume.boileau@uantwerpen.be}
\emailAdd{nelson.christensen@oca.eu}
\emailAdd{c.gowling@sussex.ac.uk}
\emailAdd{mark.hindmarsh@helsinki.fi}
\emailAdd{renate.meyer@auckland.ac.nz}

\abstract{First order phase transitions in the early universe could produce a gravitational-wave background that might be detectable by the Laser Interferometer Space Antenna (LISA). Such an observation would provide evidence for physics beyond the Standard Model. We study the ability of LISA to observe a gravitational-wave background from phase transitions in the presence of an extragalactic foreground from binary black hole mergers throughout the universe, a galactic foreground from white dwarf binaries, and LISA noise. Modelling the phase transition gravitational wave background as a 
double broken power law, we use the deviance information criterion as a detection statistic, and Fisher matrix and Markov Chain Monte Carlo methods to assess the measurement accuracy of 
the parameters of the power spectrum. While estimating all the parameters associated with the gravitational-wave backgrounds, foregrounds, and LISA noise, we find that 
LISA could detect a gravitational-wave background from phase transitions with a peak frequency of 1 mHz and normalized energy density amplitude of $\OmPeak \simeq 3 \times 10^{-11}$.
With $\OmPeak \simeq 10^{-10}$, the signal is detectable if the peak frequency is in the range $4 \times 10^{-4}$ to $9 \times 10^{-3}$ Hz, 
and the peak amplitude and frequency can be estimated to an accuracy of 10\% to 1\%.}

\begin{document}
\maketitle
\flushbottom

\setlength{\parskip}{6pt plus 1pt}

\section{Introduction}

The Laser Interferometer Space Antenna (LISA)~\cite{Audley:2017drz} will be sensitive to the millihertz gravitational wave (GW) frequency range, and will simultaneously observe signals from numerous independent sources, both astrophysical~\cite{Amaro-Seoane:2022rxf} and cosmological \cite{Christensen:2018iqi}. Of particular interest  is a search for a cosmological stochastic GW background, which could come from many different processes in the early universe,  such as cosmic strings, inflation, 
or phase transitions~\cite{Auclair:2022lcg}. Here we focus on the cosmological GW background from a first order phase transition at the electroweak scale (see e.g. \cite{Mazumdar:2018dfl,Hindmarsh:2020hop} for reviews).

Any cosmological GW background will compete with numerous foregrounds. Foregrounds from large numbers of astrophysical objects with low signal-to-noise ratio will also produce stochastic signals, which need to be separated from the cosmological signal of interest.   
A first component to consider is the foreground from double white dwarf (DWD) binaries in our galaxy \cite{Lamberts:2019nyk}, whose amplitude will be annually modulated by the orbit of the LISA constellation around the Sun \cite{Adams:2013qma}. This orbital modulation aids in the separation of the galactic foreground using the LISA data~\cite{Boileau:2021sni}.

From the LIGO-Virgo observations of binary black hole (BBH) and binary neutron star (BNS) mergers one knows that there will be a foreground created from mergers of extragalactic compact binaries (ECB) over the history of the universe; LIGO and Virgo predict a background at the level (normalized energy density) of 
$\Omecb(f) = \Aastro  ({f}/{25 \text{Hz}})^{2/3}$, 
where 
$\Aastro = 6.8^{+3.6}_{-2.2} \times 10^{-10}$~\cite{KAGRA:2021kbb}. Other studies based on the LIGO-Virgo observations predict $\Aastro \simeq 1.8 \times 10^{-9} - 2.5 \times 10^{-9}$~\cite{2019ApJ...871...97C}, and population simulations populations predict $\Aastro  \simeq 5.0 \times 10^{-9} - 2.6 \times 10^{-8}$ ~\cite{PhysRevD.103.043002}.

An important line of study is to investigate LISA's ability to separate a GW background of cosmological origin from the numerous astrophysical sources and LISA noise. In this paper we consider GWs emitted from a first order phase transition at the electroweak energy scale. Such a transition would have happened at around 10 picoseconds after the Big Bang, and generated a signal which could fall within LISA's peak sensitivity window, in the range 1 - 10 mHz.

In the Standard Model electroweak symmetry-breaking is not associated with a first order phase transition: there is instead a smooth crossover \cite{Kajantie:1996mn,Kajantie:1996qd}. 
However, in numerous extensions to the Standard model a first order phase transition is possible (for a summary see \cite{Caprini:2019egz}) turning a search for GW background into a search for physics beyond the Standard Model, which is needed for explanations for the dark matter and baryon asymmetry of the Universe.

The production of GWs during a first order phase transition occurs via the collision of bubbles of the stable phase, and the subsequently produced sound waves and turbulent flows. In a first order transition driven by thermal fluctuations, sound waves created by the expanding bubbles are the dominant source of GWs  \cite{Hindmarsh:2013xza,Hindmarsh:2015qta,Hindmarsh:2017gnf};
production by bubble collisions \cite{Cutting:2018tjt,Cutting:2020nla,Lewicki:2020azd,Lewicki:2020jiv,Lee:2021nwg,Lewicki:2022pdb} can become relevant if there is very strong supercooling \cite{Ellis:2019oqb,Ellis:2020nnr}. 

Here we assume that the sound wave component is dominant, and model the GW background component with the Sound Shell Model (SSM)~\cite{Hindmarsh:2016lnk, Hindmarsh:2019phv}.  
We include an empirical factor accounting for the kinetic energy suppression in strong transitions \cite{Cutting:2019zws}, and assume that the transition is not so strong that the modifications to the spectrum from shocks \cite{Dahl:2021wyk} and vortical turbulence \cite{Auclair:2022jod} become important.

The GW power spectrum in the SSM is determined by a few key thermodynamic parameters: the bubble nucleation temperature, $\Tn$, the phase transition strength, $\al$, the bubble wall speed, $\vw$, and the mean bubble spacing in units of the Hubble length, $r_*$. 
The sound speeds in the two phases are also important \cite{Giese:2020rtr,Giese:2020znk}.
These thermodynamic parameters are directly related to the underlying physics model. 

As it is computationally intensive to calculate a power spectrum with the full SSM it is useful to investigate LISA's sensitivity to a parametrised spectral shape that approximates the SSM and can be used for the rapid evaluations needed in Markov Chain Monte Carlo (MCMC) searches. 
Here we use a double broken power law with four spectral parameters: the peak amplitude, $\OmPeak$, the peak frequency, $\fp$, the ratio $\rb$ between the peak frequency and the break frequency and the slope between the two characteristic frequency scales $b$.  The SSM predicts that, where long-lived sound waves are the dominant source of GWs, the low frequency and high frequency spectral slopes are fixed at $9$ and $-3$. As discussed in \cite{Gowling:2021gcy}  the relationship between the spectral and thermodynamic parameters is complicated and there are numerous degeneracies. In this work we focus on LISA's ability to constrain the spectral parameters and leave the connection between thermodynamic and spectral parameters for future work.

The Fisher matrix study performed in \cite{Gowling:2021gcy} estimates LISA's sensitivity to a first order phase transition signature described by the SSM.
In that work two GW power spectra models were considered: the SSM itself, and the double broken power law fit to the SSM. Relative uncertainties for the thermodynamic and spectral parameters were calculated using Fisher analysis and a data model that takes into account LISA noise, a stationary DWD foreground and an extragalactic astrophysical background.

Another study looked at LISA's ability to detect to a general double broken power law that has the low and high frequency spectral slopes unspecified \cite{Giese:2021dnw}, instead of fixed at the SSM values. The Akaike Information Criterion was used to determine whether, in the presence of LISA instrument noise, one is able to identify the break frequencies. This work also calculated the uncertainties in spectral parameters using MCMC simulations for several example cases. The noise model included LISA instrument noise built out of one TDI channel in mock data generation, but did not include any astrophysical foregrounds.

In this present paper we investigate LISA's ability to detect a GW background from a first order phase transition, in the presence of LISA noise, the galactic foreground, and the foreground from extragalactic compact binaries. 
To do this we use the difference in the deviance information criterion (DIC) between models with and without phase transition, calculated using  MCMC methods, as a detection statistic ~\cite{Spiegelhalter2002, Spiegelhalter2014,MeyerDICreview2016}. We consider a value of $\Delta\text{DIC}>5$ to be a detection as discussed in more detail in Section~\ref{sec:FMandDIC}. In comparison to \cite{Gowling:2021gcy,Giese:2021dnw} we use the $\A\E\T$ time delay interferometery (TDI) channels to build our data model and use the $\T$ channel to constrain the LISA instrument noise in our MCMC simulations. We then perform a systematic scan over the spectral parameter space using Fisher matrix and MCMC methods to determine how well the four spectral parameters of the first order phase transition can be estimated with LISA. 

The rest of this work is organised as follows. In Section~\ref{sec:Model} we describe the GW background from a phase transition, the LISA noise model, the DWD foreground and the extragalactic compact binary foreground used in this analysis. In Section \ref{sec:Simulation} we discuss how we simulate the data. In Section \ref{sec:FMandDIC} we outline how the estimates based on the Fisher matrix and DIC are evaluated. The results and conclusions from this study are presented in Section \ref{sec:Results} and Section \ref{sec:conclusions}.

\section{Model Components}
\label{sec:Model}
The parameter estimation methods used here follow those outlined in \cite{Boileau:2021sni}, which explored parameter estimation with GW backgrounds with spectra in the form of a simple power law. 
We consider a cosmological GW background described by the double broken power law discussed in \cite{Gowling:2021gcy} which models the signal expected from a first order phase transition. In this Section we outline our model for the combined power spectrum from a first order phase transition, LISA noise, the DWD foreground and the extragalactic compact binary foreground.

With LISA's triangular geometry, the interferometric phase differences can be combined in different ways with different time delays to eliminate the laser frequency noise \cite{Tinto:2001ii,Tinto:2002de}, using the technique of time delay interferometry (TDI). This leads to the construction of three GW measurement channels known as the $\X,\Y,\Z$  TDI channels~\citep{Vallisneri_2012}. Here we assume that the GW background signal, observed in the  $\X,\Y,\Z$ channels, is stationary and uncorrelated with the stationary LISA instrument noise. Furthermore, we make the simplifying assumption that the 
instrument noise consists of two components: test mass acceleration noise and optical path length fluctuation noise; that these instrumental noises are identical in each spacecraft, 
and that arm lengths are the same so that the LISA instruments form an equilateral triangle. Under these assumptions, the cross-spectra and response functions of the $\X,\Y,\Z$ channel combinations are identical \cite{Flauger_2021}. To be conservative, we ignore the annual modulation of the galactic binary foreground.

It is possible to work with linear combinations of these channels for convenience: we choose 
the two ``noise orthogonal'' channels $A$ and $E$, and the ``null'' channel $T$ which has a reduced sensitivity to GWs, which are defined as 

\begin{equation}\label{eq:AET}
\left\{
\begin{array}{l}
\displaystyle A = \frac{1}{\sqrt{2}}(Z-X), \\[3pt]
\displaystyle E = \frac{1}{\sqrt{6}}(X-2Y+Z), \\[3pt]
\displaystyle T = \frac{1}{\sqrt{3}}(X+Y+Z).
\end{array}
\right.
\end{equation}
For ease of calculation, we use the approximation for the GW response of the $A$, $E$ and $T$ channels 
given in Ref.~\citep{Smith:2019wny}:
\begin{equation}\label{Eq:R_A_fit}
\mathcal{R}^{\rm Fit}_{A}(f) = \mathcal{R}^{\rm Fit}_{E}(f)= \frac{9}{20} |W(f)|^2 \left[1 + \left(\frac{ f}{4f_*/3}\right)^2 \right]^{-1}, 
\end{equation}

\ben\label{Eq:R_T_fit}
\mathcal{R}^{\text{Fit}}_{T} \simeq \frac{1}{4032} \left(\frac{f}{f_*}\right)^{6} |W(f)|^2 \left[ 1 + \frac{5}{16128} \left(\frac{f}{f_*}\right)^{8} \right]^{-1}
\een
where $W(f) = 1 - e^{-{2if}/{f_*}}$ with $f_*={c}/{2\pi L }$, and we take $L = 2.5\times10^9$ m as appropriate for the LISA arms. 

\subsection{Cosmological background from a first order phase transition}
\label{sec:PTfit}
The GW power spectrum from a first order cosmological phase transition is thought to be well approximated by the Sound Shell Model \cite{Hindmarsh:2019phv}, at least for transitions which are not too strong and have high enough wall speeds \cite{Cutting:2019zws}. 
In the model there are two characteristic length scales, the mean bubble separation and the sound shell thickness, which motivate a simplified description in terms of a function with two frequency scales and three power law indices - a double broken power law \cite{Hindmarsh:2019phv}.  In this work where we address a GW background from a first order phase transition, we use the double broken power law fit to the SSM put forward in \cite{Gowling:2021gcy} and shown to be a good fit over a wide range of wall speeds and transition strengths. 
In this fit, the power spectrum takes the form 
\ben{\label{Eq:omgw_dbl_brkn}}
    \OmGWPT(f;\OmPeak,\fp,r_b ,b) =\OmPeak M(f;\fp,\rb, b)
\een
where $\OmPeak$ is the peak of the power spectrum, $\fp$ is the frequency corresponding to $\OmPeak$ and  $\rb =  f_{\text{b}} /\fp$ describes the ratio between the two breaks in the spectrum. The parameter $b$ defines the spectral slope between the two breaks. The spectral shape $M(f,\fp,\rb, b)$ is a double broken power law with a spectral slope $9$ at low frequencies and $-4$ at high frequencies\footnote{In practice, the SSM's predicted high-frequency power law of $-3$ emerges only slowly, and $-4$ provides a better fit around the peak \cite{Hindmarsh:2017gnf,Hindmarsh:2019phv}.}
\ben{\label{Eq: M double_break}}
 M ( f;\fp,\rb , b ) = \left(\frac{f}{\fp}\right)^{ 9 } {\left( \frac { 1 + \rb^4 } { \rb^4 + \left(\frac{f}{\fp}\right)^4}\right)}^{(9 -b)/4}  \left( \frac { b +4 } { b + 4 - m + m \left(\frac{f}{\fp}\right) ^ { 2 } } \right) ^ { (b +4) / 2 } .
\een
Within $ M ( f;\fp, \rb , b )$, $m$ has been chosen to ensure that  for $\rb<1$ the peak occurs at $f=\fp$  and $M(\fp;\fp,\rb,b) = 1$, giving 
\ben{\label{Eq: m}}
    m = \left( 9 {\rb}^4+ b\right) / \left( {\rb}^4 +1 \right).
\een
There are regions of the spectral parameter space that lead to Eq.~\ref{Eq: M double_break} diverging. In particular, when the denominator of the final factor becomes negative, i.e.\
\ben
 b + 4 - m + m \left(\frac{f}{\fp}\right) ^ { 2 }  \le 0.
 \een
We restrict ourselves to working within the region of parameter space that is well-defined. 

Here, we outline the key thermodynamic parameters and their connection to the spectral ones. 
The first of the thermodynamic parameters is the nucleation temperature $\Tn$, which we define as the temperature corresponding to the peak of the globally-averaged bubble nucleation rate.  The Hubble rate at $\Tn$ sets the frequency scale of the GW spectrum.

The second thermodynamic parameter is the nucleation rate parameter $\be$. As discussed in \cite{Gowling:2021gcy} due to uncertainties in the calculation of $\be$, we instead consider the related quantity, the mean bubble spacing $R_*$. We note that $\be^{-1}$ is the time for the bubble wall to move a distance $R_*$ and therefore has the interpretation of the duration of the phase transition. In this work we refer to the Hubble-scaled mean bubble spacing $r_* = H_n R_*$ which contributes to  the frequency scale and amplitude of the GW power spectrum.
 
Our third key thermodynamic parameter is the phase transition strength $\al$, which we define as the ratio between the trace anomaly and the thermal energy, where the trace anomaly describes the amount of energy available to convert to shear stress energy. A stronger transition means more energy is converted to shear stress energy and a larger overall amplitude for the GW signal.

The final parameter to introduce is the wall speed $\vw$ which, along with $\al$, determines the motion of the plasma surrounding the bubble wall.  The value of the wall speed relative to speed of sound $\cs$ determines the width of the GW power spectrum, here we assume is the ultrarelativistic value $\cs = 1/\sqrt{3}$ (see \cite{Giese:2020rtr,Giese:2020znk} for other scenarios). For wall speeds close to $\cs$ the power spectra are broad and $r_b$ is small, in the alternate case the power spectra are narrow. 

To summarise, the peak amplitude is controlled by the phase transition strength, the Hubble-scaled mean bubble spacing and the bubble wall speed in rough order of efficacy from high to low. For the peak frequency all thermodynamic parameters contribute to varying degrees. It is worth noting the nucleation temperature only impacts the overall frequency scale whereas all the other thermodynamic parameters play a role in numerous spectral parameters.  The break ratio and the intermediate spectral slope are related to the phase transition strength and the wall speed parameters.
Fig.~3 of \cite{Gowling:2021gcy} demonstrates how changing the thermodynamic parameters affects the shape of the  GW power spectrum.

\subsection{LISA noise model}

We take the LISA noise model to be that given in the LISA Science Requirement Document
\citep{LISA_SR_doc} and \cite{Baker:2019nia}. The model 
assumes constant equal noise in all channels, and has only two parameters: the acceleration noise level $ \Nacc = 1.44 \times 10^{-48} \ \text{s}^{-4} \text{Hz}^{-1} $ and the optical path length fluctuation noise level $ \Npos = 3.6 \times 10^{-41} \ \text{Hz}^{-1}$. 
The noise model is then specified by the spectral density of the $X$ channel and the cross spectral density of the channel $X$ and $Y$, which are  
\begin{equation}\label{eq:lisamodel2}
\left\{
\begin{array}{l}
    N_X(f) = \left(4 P_s(f) + 8\left[1 + \cos^2\left(\frac{f}{f_*}\right)\right] P_a(f)\right)|W(f)|^2 \\
    N_{XY}(f) = -\left[2 P_s(f) + 8 P_a(f)\right]\cos\left(\frac{f}{f_*}\right)|W(f)|^2.
\end{array}
\right.
\end{equation}
We also define the functions 
\begin{equation}\label{eq6}
\left\{
\begin{array}{l}
    P_s(f) = \Npos \\
\displaystyle    P_a(f) = \frac{\Nacc}{(2 \pi f)^4}\left( 1 + \left(\frac{0.4 \text{ mHz}}{f} \right)^2 \right),
\end{array}
\right.
\end{equation} 
with $P_s(f)$ the single optical path-length fluctuation noise (which is frequency-independent) and $P_a(f)$ the single test mass acceleration noise. The noise models for the $A ET$ channel power spectral densities are given by the diagonalization of the covariance matrix of the $XYZ$ channels (see e.g.~\cite{Smith:2019wny}). 
The diagonal entries are then 
\begin{equation}
\begin{array}{l}\label{eq:lisamodel}
    N_A(f) = N_E(f) = N_X(f) - N_{XY}(f), \\
    N_T(f) = N_X(f) + 2 N_{XY}(f), 
\end{array}
\end{equation}
using the assumption that the correlation noise is the same for all interferometers. 

Rather than comparing the detector response to a stochastic GW signal to the noise, it is more convenient to introduce noise spectral densities $S_A(f)$ and $S_E(f)$ by dividing by the GW response function, 
\begin{equation}
S_A(f)=S_E(f)= \frac{N_A(f)}{\mathcal{R}^{\rm Fit}_{A,E}(f)},
\end{equation}
where $\mathcal{R}^{\rm Fit}_{A,E}(f)$ is given by Eq.~\ref{Eq:R_A_fit}. For completeness the noise spectral density in the $T$ channel is 
\ben
S_{T}(f) = \frac{N_T(f)}{\mathcal{R}^{\rm Fit}_{T}(f)},
\een
where $\mathcal{R}^{\rm Fit}_{T}(f)$  is given in Eq.~\ref{Eq:R_T_fit}.
From the noise spectral densities, the equivalent energy spectral density is given by 
\begin{equation}
\Omega_A(f) = \Omega_E(f) = S_A(f)\frac{4\pi^2f^3}{3H_0^2}.
\end{equation}
These power spectra have the interpretation as an isotropic GW signal 
which would have unit signal-to-noise ratio at every frequency.

\subsection{Double white dwarf foreground}

A foreground from DWD binaries in our galaxy \cite{Lamberts:2019nyk,Nelemans_2001, PhysRevD.76.083006, Ruiter_2010, 2014PhRvD..89b2001A, 2017PASA...34...58E, Hernandez_2020, PhysRevD.64.121501}, will be observed as a modulated waveform due to LISA's orbit around the Sun. The large majority of the DWD will not be resolved, and the superposition of all GWs received by LISA constitutes the galactic foreground (see Eq.~\ref{eq:waveform}). We assume that the waveform of each binary can be modelled as a pseudo-monochromatic signal. Thus, we can build the superposition of their GW signals $s(t)$:
\begin{equation}\label{eq:waveform}
s(t) = \sum_{i=1}^{N} \sum_{P= +,\times } h_{A,i}(f_{orb,i},M_{1i},M_{2i},{\bf r}_i,t) \times F_P(\theta,\phi,t) \textbf{D}(\theta,\phi,f)_P:\textbf{e}_{P}
\end{equation}
with $i$ labelling the binaries. The masses of the two stars are $M_{1i}$ for the larger mass and $M_{2_i}$ for the smaller; the orbital frequency of the binary is $f_{\text{orb}_i}$; the Cartesian position in the Galaxy ${\bf r}_i$ and the position in the sky $\theta,\phi$; $F_A$ is the beam pattern function for the polarization $A = +,\times$, $\textbf{h}_{A,i} = h_{A,i} \textbf{e}_{A}$ the tensor of the amplitude of the GW; $\textbf{D}$ the one-arm detector tensor; and $\textbf{h}_{A,i}$ the dimensionless GW amplitude. 
An initial description has been done with resolved sources to provide the modulation of the foreground from the LISA orbit~\cite{Adams:2013qma}.

The dimensionless energy spectral density of the DWD foreground can be approximated by a broken power law. The broken power law model  for the galactic foreground from Lambert {\it et al.} used in Boileau {\it et al.}~\cite{Boileau:2021gbr} is given by 
\begin{equation}\label{Eq:SGWBDWD}
\OmDWD(f) = \frac{A_1 \left({f}/{f_*}\right)^{\alpha_1}}{1 + A_2 \left({f}/{f_*}\right)^{\alpha_2}} 
\end{equation}
with $\alpha_1-\alpha_2 = \alpha \simeq	 2/3$ at low-frequencies and the frequency reference $f_* = {c}/{2\pi L}$. 
The spectral shape of the DWD foreground is a broken power law because at high-frequencies ($\simeq 0.1$ Hz) the number of DWDs decreases due to the physical limitation from the respective radii of the two white dwarfs in each binary. 

A different model of the DWD foreground is used in Ref.~\cite{Robson_2019}.  Here, the 
frequency break is much higher, and the signal is approximated as a simple power law~\ over the LISA frequency range.  With this model, it is important to account for the resolved binaries, which are then removed, leaving behind a confusion noise $S_c(f)$ from the unresolved binaries. For a LISA mission duration of 4 years, the confusion noise from unresolved DWDs is approximated by:
\begin{equation}\label{Eq:Cornish_DWD}
S_c(f) = Af_\text{Hz}^{7/3}e^{-f_\text{Hz}^{\alpha}+ \beta f_\text{Hz}\sin(\kappa f_\text{Hz})} \left[1 + \tanh (\gamma (f_k - f_\text{Hz})) \right]
\end{equation}    
with $f_\text{Hz} = f/(1\; \text{Hz})$, 
$\alpha = 0.138$, $\beta = -221$, $\gamma=1680$ and $f_k = 0.00113$~\footnote{We note that the models for galactic confusion noise continue to be improved~\cite{Karnesis:2021tsh}, and we will incorporate these advances in future work.}.
The corresponding dimensionless energy spectral density is then 
\begin{equation}
\Omega_c(f) = S_c(f)\frac{4\pi^2f^3}{3H_0^2} ~ .
\end{equation}

\subsection{Extragalactic compact binary foreground}

A background from compact binaries consisting of black holes and neutron stars in other galaxies is expected. As this background has not yet been detected at ground-based GW observatories for this work we estimate its amplitude from the LIGO-Virgo observations, as outlined in ~\cite{Chen:2018rzo}.
Our model for this energy spectral density is a power law
\ben\label{Eq:OmAstro}
\Omecb(f) = \Aastro \left(\frac{f}{f_{\rm ref}}\right)^{\alAstro}.
\een
In our simulations we inject an extragalactic compact binary foreground with a spectral slope $\alAstro = \frac{2}{3}$  and $f_{\rm ref} = 25$ Hz as the reference frequency~\cite{Farmer:2003pa}.  For the amplitude $\Aastro$ we use the upper value from the LIGO-Virgo O2 limit distribution, $\Aastro =\Omecb(25{\rm \ Hz}) = 2.15 \times10^{-9}$~\cite{LIGOScientific:2019vic}.

\subsection{Model illustration}
\begin{figure}
    \centering
    \includegraphics[width=\linewidth]{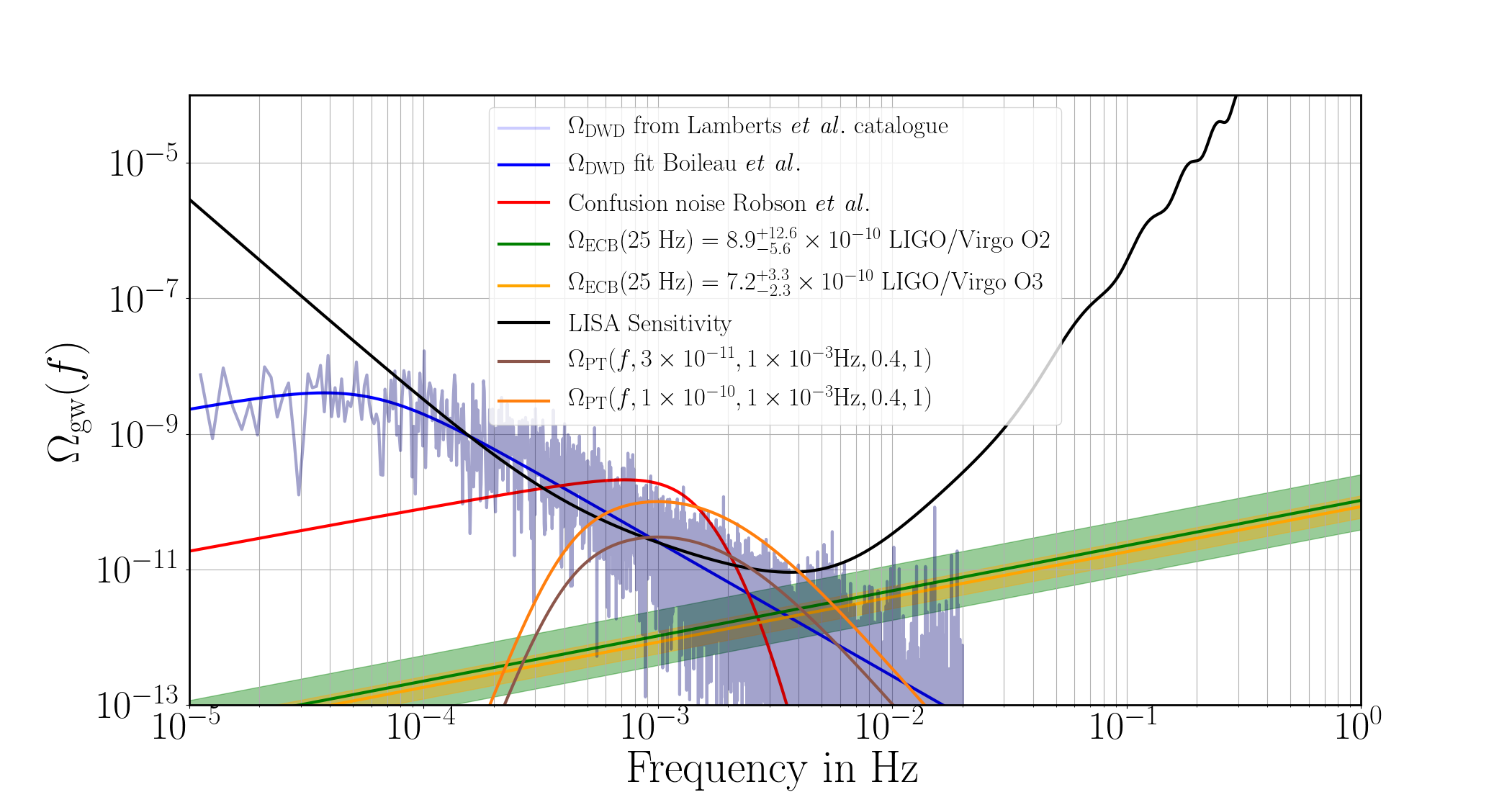}

    \caption{The LISA sensitivity curve \cite{LISA_SR_doc,Babak:2021mhe}, black line, in terms of the dimensionless energy spectral density  $\OmGW(f)$.  The DWD foreground models are also presented. The light and dark blue lines are respectively the Galactic foreground from the Lamberts \textit{et al.} catalogue \citep{Lamberts:2019nyk} and the analytic galactic foreground fit of Boileau {\it et al.}~\citep{Boileau:2021gbr}. The red line is the Galactic confusion noise from Robson \textit{et al.}~\citep{Robson_2019}. The green line is the estimated extragalactic compact binary foreground from the LIGO-Virgo 02 data~\cite{Chen:2018rzo}, while the yellow curve is estimation from the LIGO-Virgo 03 data~\cite{KAGRA:2021kbb}. The pink and orange lines are PT broken power law models with $\OmPeak  = 3\times 10^{-11}$ and $\OmPeak = 1 \times 10^{-10}$.}
    \label{fig:OmegaGW}
\end{figure}

The various contributions to the dimensionless energy density power spectrum $\OmGW(f)$ in the LISA observational band are displayed in Fig.~\ref{fig:OmegaGW}.  The black line is the LISA design sensitivity~\cite{Babak:2021mhe}.  We display three models for the galactic foreground: the Lamberts et al. catalogue DWD (light blue line)~\citep{Lamberts:2019nyk}, the Boileau et al. broken power law (dark blue)~\citep{Boileau:2021gbr}, and the galactic confusion noise of Robson et al. (red line)~\cite{Cornish:2018dyw}. The green line is the LIGO-Virgo O2 observations $\Omecb(25{\rm \ Hz}) = 8.9^{+12.6}_{-5.6} \times10^{-10}$ \cite{LIGOScientific:2019vic}, and the yellow is the LIGO-Virgo O3 observation measurement $\Omecb(25{\rm \ Hz}) = 7.2^{+3.3}_{-2.3} \times10^{-10}$ \cite{KAGRA:2021kbb}. The pink and orange lines are respectively the dimensionless energy power spectrum of the PT for $\OmPeak = 3 \times 10^{-11}$ and $\OmPeak = 1 \times 10^{-10}$. The two curves are given by Eq.~\ref{Eq:omgw_dbl_brkn} with $\fp = 1 {\rm mHz}$, $r_b = 0.4$ and $b=1$.

\subsection{Simulation}
\label{sec:Simulation}

To simulate the data in the frequency domain, we use the fractional energy density power spectrum of a first order phase transition $\OmPT (f)$ (see Section~\ref{sec:PTfit}). 
The phase transition model is parametrised by four parameters; the peak power $\OmPeak$, the peak frequency $\fp$, the break ratio $r_b$ and the intermediate power law $b$ as described in Eq.~\ref{Eq:omgw_dbl_brkn}. We also include the DWD foreground  $\OmDWD (f)$ (see Eq.~\ref{Eq:SGWBDWD}), and a power law for the extragalactic compact binary foreground, $\Omecb (f)$
(see Eq.~\ref{Eq:OmAstro}). The total GW background is the sum
\begin{equation}\label{Eq:OmGWmodel}
\OmGW(f) = \OmDWD(f;A_1,A_2,\alOne, \alTwo) + \Omecb(f; \Aastro, \alAstro)  + \OmGWPT(f; \OmPeak, \fp, r_b, b).
\end{equation} 
The total GW power spectrum model then has ten independent parameters given by\linebreak
$\theta_{\rm gw} = (\OmPeak, f_p, r_b, b, A_1, \alpha_1, A_2, \alpha_2, \Aastro, \alAstro)$.
We also model and simulate the LISA noise with the two magnitude parameters $\theta_{LISA} = (\Nacc, \Npos)$.
The parameter vector $\theta = \theta_{\rm gw} \cup \theta_{LISA}$ used in this study has 12 components from the GW background model and the LISA noise model. 
For the MCMC runs, data are simulated in the frequency domain, with a linear frequency vector of 100,000 points in the frequency band $[1\times 10^{-5}, 1]$ Hz. 

\begin{table}[h!]
\centering
\begin{tabularx}{0.5\textwidth} { 
  | >{\raggedright\arraybackslash}X 
  | >{\centering\arraybackslash}X 
  | >{\raggedleft\arraybackslash}X | }
 \hline
 $\Nacc$ & $1.44 \times 10^{-48} \ \text{s}^{-4} \text{Hz}^{-1}$ \\
 \hline
 $\Npos$  & $3.6 \times 10^{-41} \ \text{Hz}^{-1}$  \\
\hline
 $\AOne$  & $7.44 \times 10^{-14}$  \\
\hline
 $\ATwo$  & $ 2.96 \times 10^{-7}$  \\
\hline
 $\alOne$  & $-1.98$  \\
\hline
 $\alTwo$  & $-2.6$  \\
\hline
 $\alAstro$  & $2/3$  \\
 \hline
  $\Aastro$  & $2.15 \times10^{-9}$  \\
\hline
\end{tabularx}
\caption{ Parameter values used in the data simulation described by Eq.~\ref{Eq:OmGWmodel}, excluding the four phase transition parameters. }
\label{table:1}
\end{table}

The data are produced in the frequency domain by generating $N=10^5$ independent 3-component Gaussian random vectors with mean zero and covariance matrix given by the spectral density matrix $\mathcal{C}_{IJ}(\theta,f_k)$ defined in Eq.~\ref{eq:Cor} for equally spaced $f_k$   between $5\times10^{-6}$ Hz and  Nyquist frequency ${1}/{2 \Delta t}=  0.5$ Hz, with a frequency resolution of $5\times10^{-6}$ Hz and a time resolution of $\Delta t= 1$ s. $\mathcal{C}_{IJ}(\theta,f_k)$ corresponds to the noise energy spectral density matrix, rescaled by the factor of ${N}/{N_{T_\text{obs}}}$ where  $N_{T_\text{obs}}$ denotes the total number of Fourier frequencies for a time series of 4 years sampled at 1 Hz. This corresponds to segmenting a 4 year long data set sampled at 1 Hz into segments of $1.16$ days and averaging over the spectra of the individual segments.

The astrophysical background is derived from the non-continuous compact binary merger signals, known as a "popcorn" background~\cite{PhysRevD.92.063002}. This background is non Gaussian. If the merger signals are long in duration and large in their rate, the signals overlap and produce a more continuous signal. This continuous signal approaches Gaussianity via the central limit theorem. It has previously been demonstrated that even with such a non Gaussian background the standard GW background searches still can detect the signal~\cite{PhysRevD.92.063002}. For our data in this present study we have assumed ideal stationary noise, no glitches, the absence of instrumental lines and gaps. The galactic foreground is modulated in amplitude~\cite{Boileau:2021sni} because of the LISA constellation orbit around the Sun. We have averaged over small segments of time (1.16 days); it has previously demonstrated that the modulation can be assumed as constant within small segments~\cite{Boileau:2021sni}. The phase transition GW background we considered is generated from the overlap of many sound waves, and can be considered as Gaussian.

For the different MCMC analyses we simulate numerous different sets of data with an independent variation on the phase transition parameters ($\OmPeak, \fp , r_b , b$).
  The goal is to estimate the impact each phase transition parameter has on the overall observability and parameter estimation. 
  We use the same parameter estimation methods as in Boileau  {\it et al.} ~\citep{Boileau:2021gbr} treating all parameters of  the GW background $\theta_\text{gw}$ and LISA noise  $\theta_{LISA}$ as unknown and estimating these simultaneously.

\subsection{Fisher information and Deviance Information Criterion}
\label{sec:FMandDIC}

The likelihood function with the data ${\cal D} = (d_A,d_E,d_T)$ uses the Fourier transform vectors for the channels $AET$. The data is in the frequency domain, given the model parameters $\theta$, and gives the likelihood

\begin{equation}\label{eq:lik}
    \begin{aligned}
        \mathcal{L}({\cal D}|\theta) &= \prod_{k=0}^N \frac{1}{\sqrt{\det\left(2\pi \mathcal{C}(\theta,f_k)\right)} } e^{-\frac{1}{2}{\cal D}^{*T}_k\mathcal{C}^{-1}(\theta,f_k){\cal D}_k} ~ ,
        \\
    \end{aligned}
\end{equation}
where the product is over the  of Fourier frequencies $f_k$, and ${\cal C}$ denotes the cross powers  spectral covariance matrix with components 
\begin{equation}\label{eq:Cor}
    \mathcal{C}(\theta,f) = \frac{3H_0^2}{4\pi^2 f^3}
     \left(
     \begin{array}{ccc}
      \left(\Omega_A(f) + \OmGW(f)\right){\cal R}_{A,E}^\text{Fit} & 0 & 0  \\
      0 & \left(\Omega_E(f) + \OmGW(f)\right){\cal R}_{A,E}^\text{Fit} & 0 \\
      0 & 0 &  \Omega_T(f){\cal R}_{T}^\text{Fit}  \\
     \end{array}
     \right).
 \end{equation} 
The dimensionless energy spectral density of the GW signal contributes equally to 
 channels $I=[A,E]$, and we neglect the response of the $T$ channel to GWs. In this work we use a simple model for the $T$ channel that contains, by definition, no GW signal. 
We understand that the real data channels will not be as simple and leave the inclusion of second generation TDI channels for future work. Here, the focus is on whether we can separate the PT GW background from the instrument noise and astrophysical foregrounds. We note too that a recent study of the triangular configuration for non-equal detector noise and non-equal noise correlations changes the properties of the $A$ and $E$ channels~\cite{https://doi.org/10.48550/arxiv.2205.00416}.
Using the channels $A$ and $E$ without channel $T$ would increase the uncertainties of the parameter estimation, especially for values near the level of detectability ($\OmPeak \sim 3 \times 10^{-11}$, presented below). Without the use of the $T$ channel there would be additional confusion between the LISA noise and the GW signals. Consequently, the limit of detectability of the GW background would increase.
To keep the notation compact we have omitted explicit notation for the sum over frequency bins $k$.


The Fisher information matrix $F_{ab}$ is used to estimate the parameters with the uncertainty $\sqrt{F_{aa}^{-1}}$ of the Fisher information  (see Eq.~\ref{eq:Fisher}) from the likelihood (see Eq.~\ref{eq:lik}): 
\begin{equation} \label{eq:Fisher}
F_{ab} = \frac{1}{2} \sum_{I=A,E,T} \sum_{k=0}^N T_\text{obs}\Delta f_k
\frac{\partial \ln \mathcal{C}_{II}(f_k)} {\partial \theta_a}
\frac{\partial \ln \mathcal{C}_{II}(f_k)} {\partial \theta_b}.
\end{equation} 
Here, $T_\text{obs}$ is the time duration of observations for the LISA mission, assumed to be 4 years, and $\Delta f_k = f_k-f_{k-1}$. 
To reduce the number of calculations, we can assume that parameters from different sources are independent and that $\theta$ can be grouped into LISA noise, extra galactic compact binary, DWD and phase transition  parameters as
\ben
\th = (\Nacc, \Npos, \Aastro, \alAstro, A_1, \alpha_1, A_2, \alpha_2, \OmPeak, \fp, r_b, b). 
\een   
The Fisher information matrix is then a block diagonal matrix
\begin{equation}
   F_{ab}(\theta)= \left( 
    \begin{array}{c|c|c|c} 
      \Gamma_{LISA} & 0 &0&0 \\ 
      \hline 
      0 & \Gamma_{\rm ECB} &0&0 \\
      \hline 
      0 & 0 &\Gamma_{\rm DWD}&0 \\
      \hline 
      0 & 0 &0&\Gamma_{\rm PT} \\
    \end{array} 
    \right) ,
\end{equation}
with respectively the Fisher information matrix of the LISA noise $ \Gamma_{LISA}$, the extragalactic compact binary background  $\Gamma_{\rm ECB}$, the DWD foreground $\Gamma_{\rm DWD}$ and the phase transition background $\Gamma_{\rm PT}$. 
Thus, the inverse of the Fisher matrix is 
\begin{equation}
   F_{ab}^{-1}(\theta)= \left( 
    \begin{array}{c|c|c|c} 
      \Gamma_{LISA}^{-1} & 0 &0&0 \\[2pt] 
      \hline 
      0 & \Gamma_{\rm ECB}^{-1} &0&0 \\[2pt]
      \hline 
      0 & 0 &\Gamma_{\rm DWD}^{-1}&0 \\[2pt]
      \hline 
      0 & 0 &0&\Gamma_{\rm PT}^{-1} \\[2pt]
    \end{array} 
    \right) 
\end{equation}

The uncertainty in parameter $\theta_a$ is 
estimated as $\sigma_a = \sqrt{F_{aa}^{-1}}$. 
In the following, we will study only the sub-matrix of the phase transition parameters $\Gamma_{\rm PT}^{-1}$. As a cross check, we also use MCMC methods to estimate the posterior distribution of the signal parameters, $p(\theta|{\cal D}) \propto p(\theta) \mathcal{L}({\cal D}|\theta)$.

The Fisher information matrix is obtained  by calculating the second order partial derivatives of the log-likelihood function with respect to unknown parameters.
There will be non-zero off-diagonal entries of the Fisher information matrix. However, for the sake of faster computation, 
 we assume a block diagonal matrix where parameters within each block $\Gamma_{LISA}$, $\Gamma_{ECB}$, $\Gamma_{DWD}$ and $\Gamma_{PT}$,  are potentially dependent but parameters from different blocks are independent. We refer to our MCMC results in Section~\ref{sec:Results}, where indeed the parameters from different blocks have negligible posterior correlation; see Figure~\ref{fig:cornerplot} in Appendix~\ref{app:MCMC_plot}.  

 The posterior distribution of the parameter vector $\theta$ is obtained by combining the likelihood in Eq.~\ref{eq:lik} with independent general priors for each of the parameters. We specify independent Gaussian priors
  \ben\label{Eq:gaussian_prior}
  p(\theta) = \prod_i \exp\left(-\frac{(\theta_i-\mu_i)^2}{2\sigma_i^2}\right)
  \een
  for the GW background, DWD and LISA noise parameters where $\mu_i$ is the true value and $\sigma_i = 1$.
    For example, for the parameter $\OmPeak$,  we sample on $\log(\OmPeak)$ with a Gaussian prior centred on the ``true'' value $\log(\OmPeak)$, with $\log(\sigma_{\OmPeak}) = 1$. We use log-parameters for $\Nacc$, $\Npos$, $\Aastro$, $A_1$, $A_2$, $\OmPeak$, and $f_p$ and sample directly with $\alOne, \alTwo$, $\alAstro$, $\rb$ and $b$.
We use the MCMC algorithm of~\cite{doi:10.1198/jcgs.2009.06134}.  
This is an adaptive Metropolis-Hastings algorithm with a proposal distribution:
\begin{equation}
Q_n(\theta)= (1-\beta){\cal N}(\theta,(2.28)^2 \Sigma_n / d ) + \beta {\cal N}(\theta,(0.1)^2 I_d/d) ~ ,
\end{equation}
where $\Sigma_n$ is the current empirical estimate of the covariance matrix of the parameter vector $\theta$ (based on the previous MCMC samples), $\beta = 0.01$, $d$ the number of parameters, $I_d$ the identity matrix and ${\cal N}$ the multi-normal distribution.

The ultimate aim is to study whether the model that includes phase transitions provides a substantially better fit than a model without phase transitions where both models include the LISA noise, DWD foreground and the compact binary produced GW background. Within the Bayesian framework, this model comparison could be performed by computing Bayes factors. However, when using improper priors or even very vague priors, these are not well defined.
The sensitivity of the Bayes factor to the choice of increasingly diffuse priors is well known and often referred to as {\em Lindley's paradox} \cite{LindleyD.V.1977Apif, ShaferGlenn1982LP}. It is illustrated for instance in \cite{MAURI2016570} for an example of a Gaussian likelihood with unknown mean $\theta$ and unknown variance $\sigma^2$ where a Normal(0,$\tau^2)$ prior is put on the variance parameter $\sigma^2$. With increasing $\tau^2$,  the marginal likelihood of the null model $\theta=\theta_0$ and that of the alternative  will converge to 1 and zero, respectively, no matter the value of the data. Thus the Bayes factor  for comparing the null to the alternative model will go to infinity even if the observed data value is far away from $\theta_0$.
 Therefore, we use the DIC~\cite{Spiegelhalter2002,Spiegelhalter2014} which can be regarded as the Bayesian analogue of the AIC/BIC and a Bayes factor approximation, and can be used even if improper priors have been specified. The DIC is a very popular choice for practical model comparison as it is easy to compute when a MCMC sample of the posterior distribution is available~\cite{MeyerDICreview2016}. 

The DIC combines a model fit statistic with a term that penalizes the model complexity. It is based on the deviance $D(\theta)$ defined as $D(\theta) = -2 \log {\cal L({\cal D} |\theta)}$, and evaluated at the posterior mean $\bar{\theta}$ of $\theta$ (the average of the posterior samples from the MCMC). The penalty term is given by  $p_D = \bar{D} -D(\bar{\theta})$ where $ \bar{D}$ denotes  the posterior mean of the deviance. The DIC is given by 
\begin{equation}
\text{DIC} = D(\overline{\theta}) + 2 p_D. 
\end{equation} 
We calculate the difference in DIC for the models with a phase transition and without. We follow the general rule of thumb  that a difference in the DIC of $\Delta \text{DIC} > 5$ there is substantial evidence for the model with a phase transition, and we have strong and decisive evidence for $\Delta \text{DIC} > 10$ \cite{LunnDavid2013TBb}. In this study we use the level of $\Delta \text{DIC} > 5$ as the threshold for detectability.

\section{Results}
\label{sec:Results}
\subsection{DIC results}
\begin{figure}
\begin{subfigure}{.5\textwidth}
  \centering
  \includegraphics[width=\linewidth]{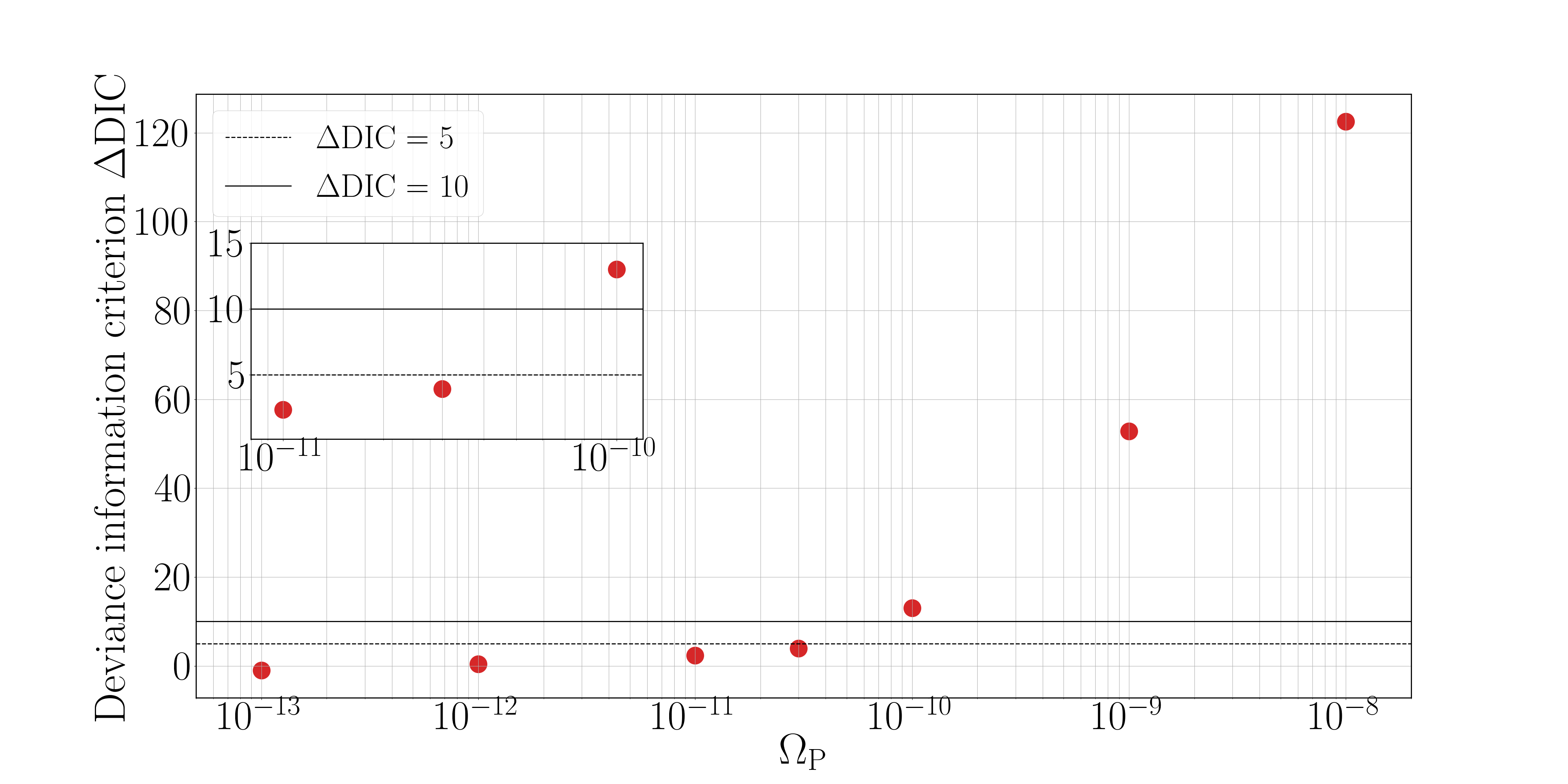}
  \caption{fixed: $\fp = 1$ mHz, $r_b = 0.4$, $b = 1$}
  \label{fig:DICOmega}
\end{subfigure}%
\begin{subfigure}{.5\textwidth}
  \centering
  \includegraphics[width=\linewidth]{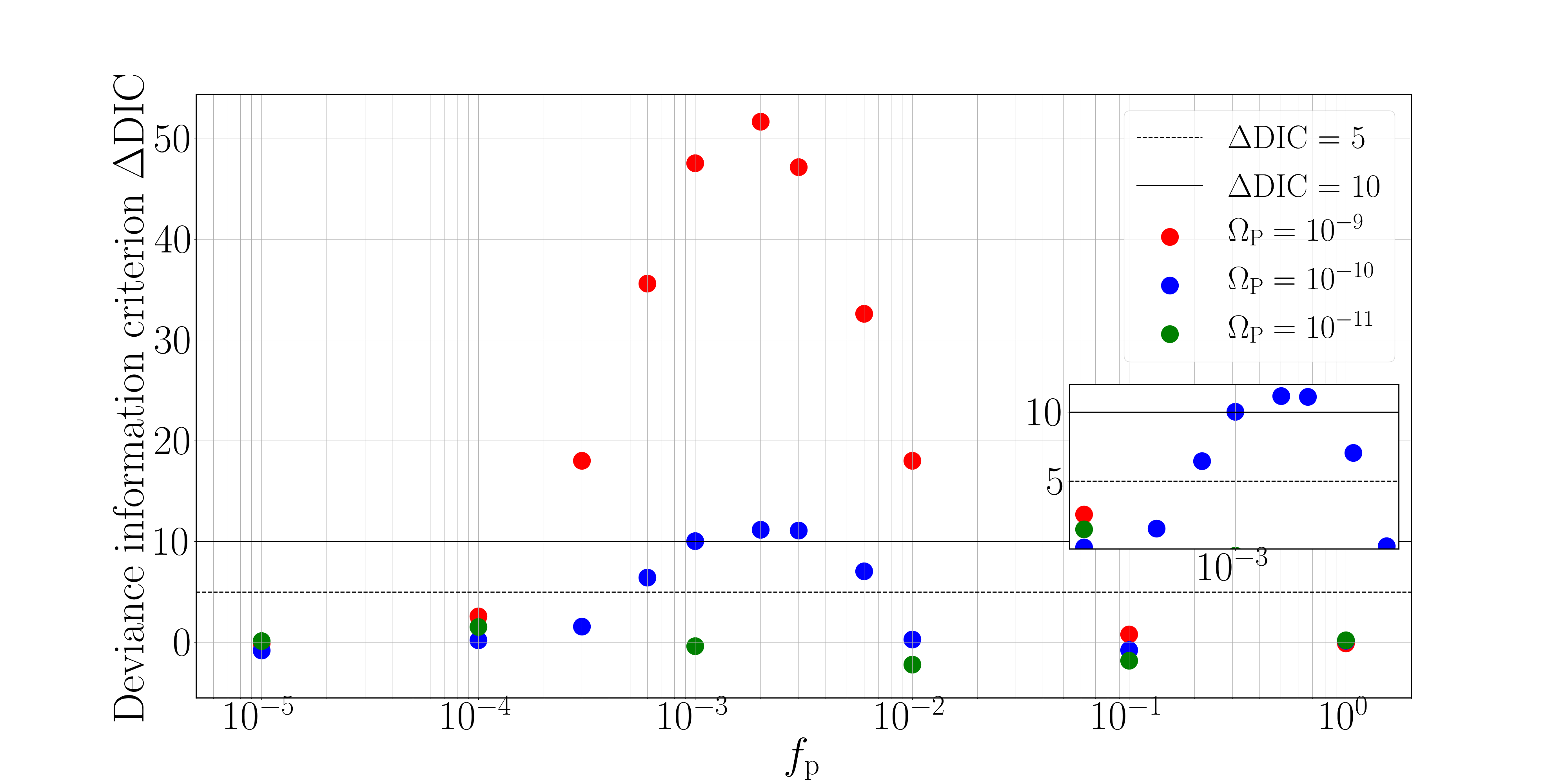}
  \caption{fixed: $r_b = 0.4$, $b = 1$}
  \label{fig:DICfp}
\end{subfigure}
\caption{Fig.~\ref{fig:DICOmega} shows the changes in the deviance information criterion ($\Delta$DIC) as the peak amplitude $\OmPeak$ is varied,  when the peak frequency $\fp = 1 \times 10^{-3}$ Hz. Fig.~\ref{fig:DICfp} shows the changes in the $\Delta$DIC as the peak frequency $\fp$ is varied, for three values of peak amplitude $\OmPeak = $ $1 \times 10^{-9}$(red), $1 \times 10^{-10}$ (blue) and $ 1 \times 10^{-11}$ (green).  In both cases the break ratio and the intermediate slope are fixed to $r_b =0.4$, and $b =1$.}
\label{fig:DIC1}
\end{figure}

We use the DIC to investigate LISA's sensitivity to a GW background from a first order phase transition in the presence of foregrounds from DWDs in the galaxy and extragalactic compact binaries. We explore how this sensitivity varies as a function of the parameters 
of the fit to the phase transition signal. 
The peak amplitude $\OmPeak$ and the peak frequency $\fp$ are the parameters that play the greatest role in determining whether one can distinguish between models with or without a phase transition signature.  In Fig.~\ref{fig:DICOmega} and Fig.~\ref{fig:DICfp} we show $\Delta \text{DIC}$ as a function of these 
 parameters.  
 For a signal peaking at $1$ mHz, which is the most favourable frequency for detection by LISA, 
 $\Delta \text{DIC}$ is above 5 for peak amplitudes $\OmPeak \gtrsim 3 \times 10^{-11}$, 
 and above 10 for peak amplitudes $\OmPeak \gtrsim 1 \times 10^{-10}$.
A signal of with magnitude $\OmPeak = 1\times10^{-10} $ has $\Delta \text{DIC} > 5 $ over a 
band from $3 \times 10^{-4}$ to $10^{-2}$ Hz, where we use the level of $\Delta \text{DIC} > 5$ as the threshold for detectability for the model with a phase transition.

\begin{figure}
\begin{subfigure}{.5\textwidth}
  \centering
  \includegraphics[width=\linewidth]{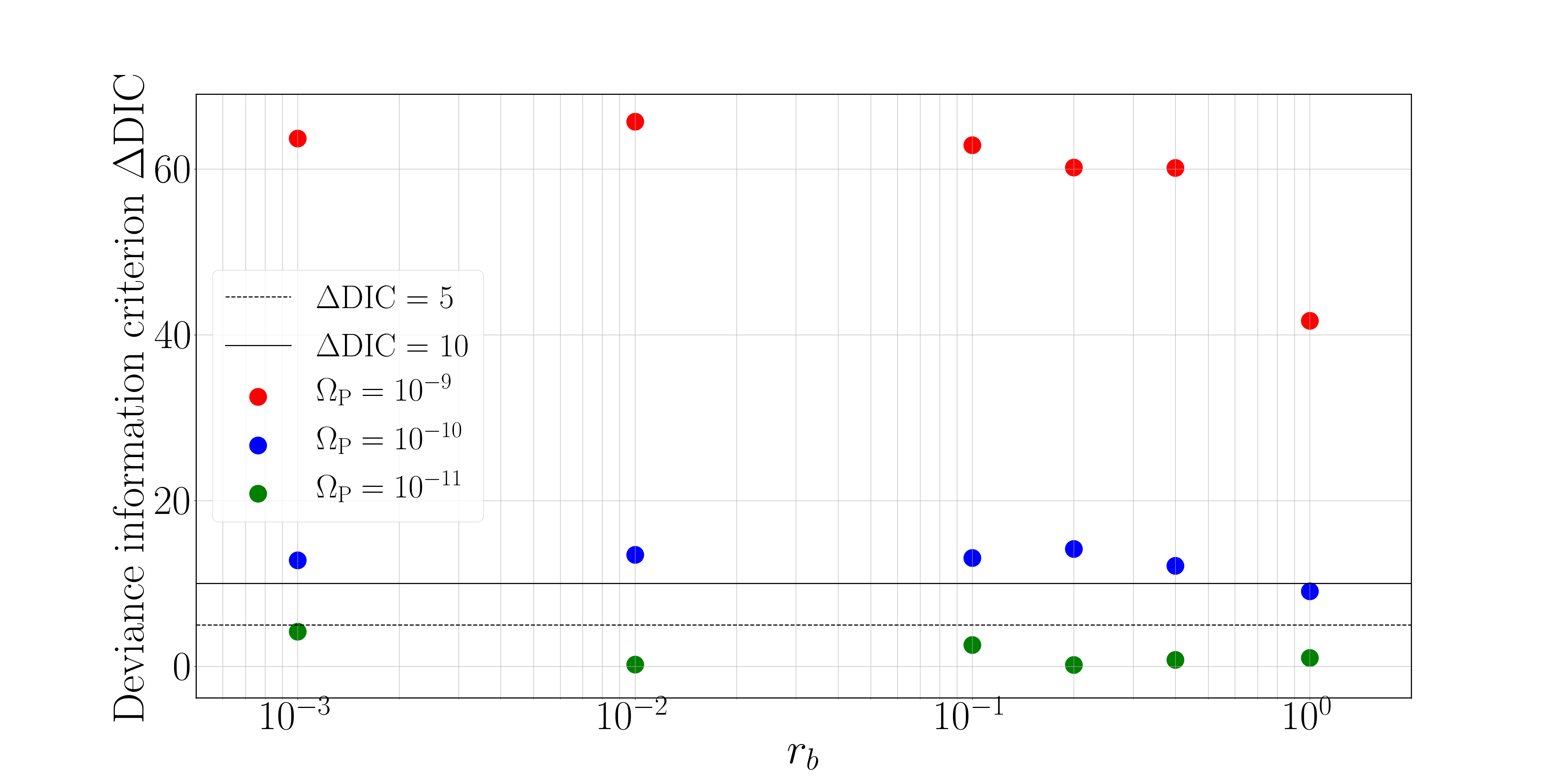}
  \caption{fixed: $\fp = 1$ mHz, $b = 1$}
  \label{fig:DICrb}
\end{subfigure}%
\begin{subfigure}{.5\textwidth}
  \centering
  \includegraphics[width=\linewidth]{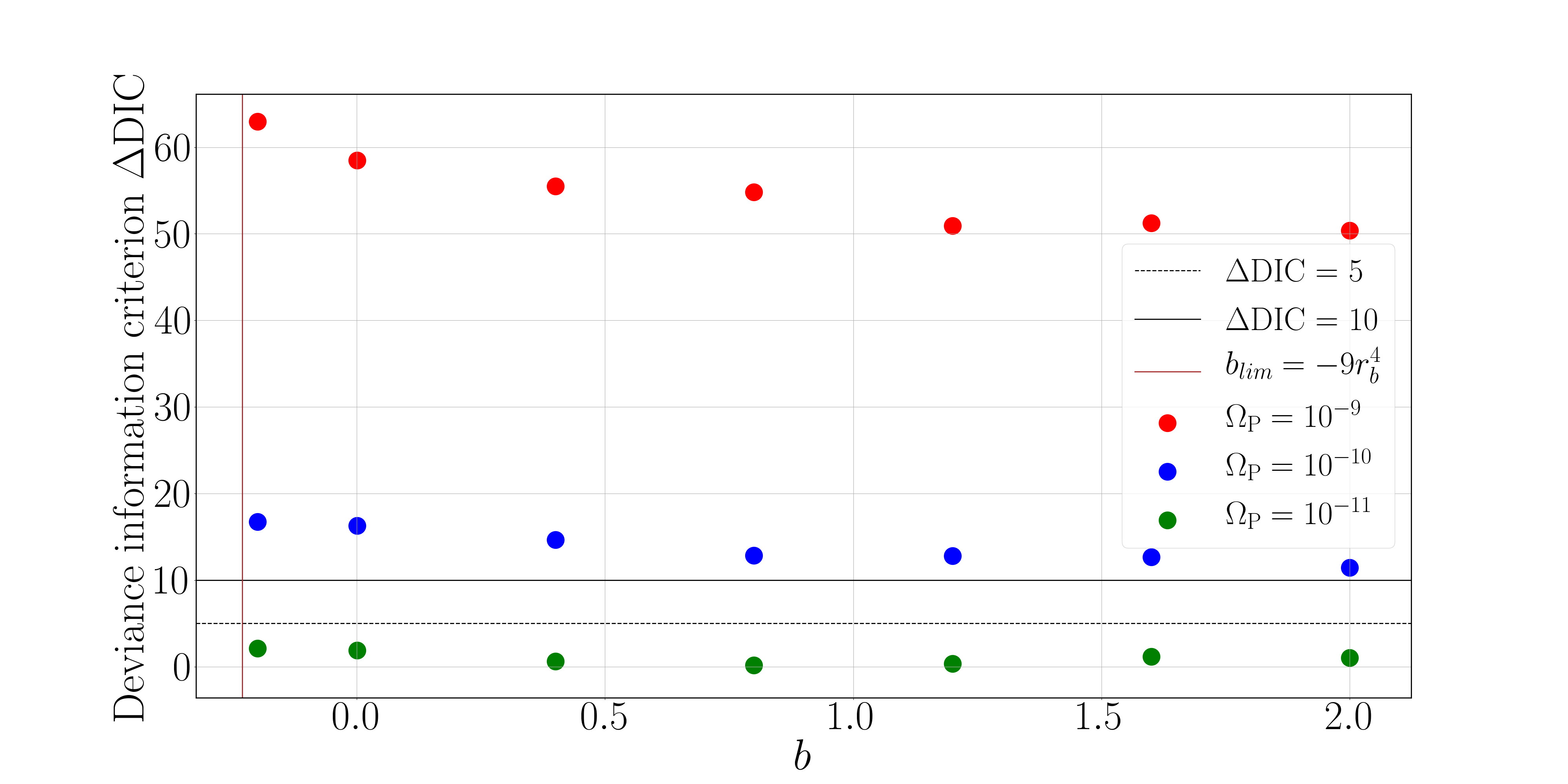}
  \caption{ fixed: $\fp = 1 $ mHz, $r_b = 0.4$}
  \label{fig:DICb}
\end{subfigure}
\caption{Fig.~\ref{fig:DICrb} displays the changes in the deviance information criterion ($\Delta$DIC) as the break ratio $\rb$ is varied and the intermediate slope $b =1$. Fig.~\ref{fig:DICb} shows the changes in $\Delta$DIC as the intermediate slope  $b$ is varied; here $\rb = 0.4$.  In both cases we considered three values of peak amplitude $\OmPeak $ $1 \times 10^{-9}$ (red), $1 \times 10^{-10}$ (blue) and $ 1 \times 10^{-11}$ (green) and the peak frequency $\fp = 1$ mHz.}
\label{fig:DIC2}
\end{figure}

In Fig.~\ref{fig:DICrb} and Fig.~\ref{fig:DICb} we see that varying the break ratio $r_b$ and the intermediate slope $b$ have little impact on the overall observational prospects of the phase transition signal. Instead we again see the importance of the peak amplitude parameter. 

 As discussed in Sec.~\ref{sec:PTfit} the relationship between the spectral parameters and the thermodynamic parameters of a first order phase transition is complicated, which makes it challenging to say anything concrete about LISA's sensitivity to the thermodynamic parameters from these results alone. The DIC analysis has shown the spectral parameters with the biggest impact on resolving a PT signature are the peak amplitude which relates to ($\al,\vw, r_*$) and the peak frequency which all thermodynamic parameters contribute to. For a more quantitative description of how the uncertainties in the spectral parameters translate into uncertainties in thermodynamic parameters see Gowling et al. \cite{Gowling:2022pzb}.

\subsection{Fisher matrix and MCMC comparison }

Here, we compare the uncertainties in the measurements of spectral parameters when calculated with the Fisher matrix \citep{Gowling:2021gcy} to those computed with MCMC simulations. 
In the Fisher method,  the relative uncertainties  are calculated using the Fisher matrix $F_{ab}$, as outlined in Eq.~\ref{eq:Fisher}, and are given by $\sqrt{F_{aa}^{-1}}$. 
The Fisher matrix is evaluated with 200 points in a log-frequency band between $[1 \times 10^{-5},1]$ Hz according to Eq.~4.3 in \citep{Gowling:2021gcy}.

\begin{figure}
\begin{subfigure}{.5\textwidth}
  \centering
  \includegraphics[width=\linewidth]{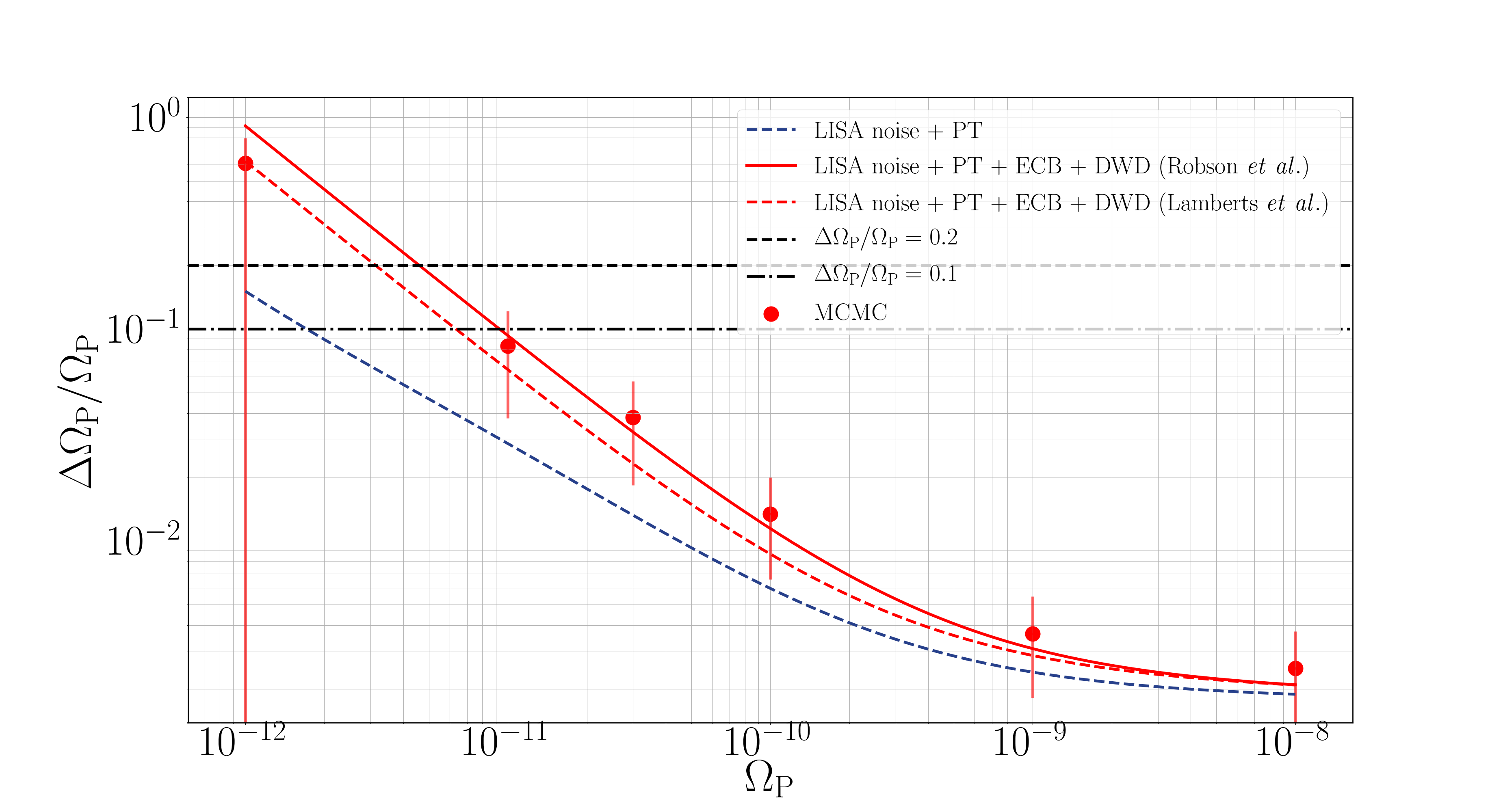}
  \caption{Fisher $\OmPeak$}
  \label{fig:FisherOmega}
\end{subfigure}%
\begin{subfigure}{.5\textwidth}
  \centering
  \includegraphics[width=\linewidth]{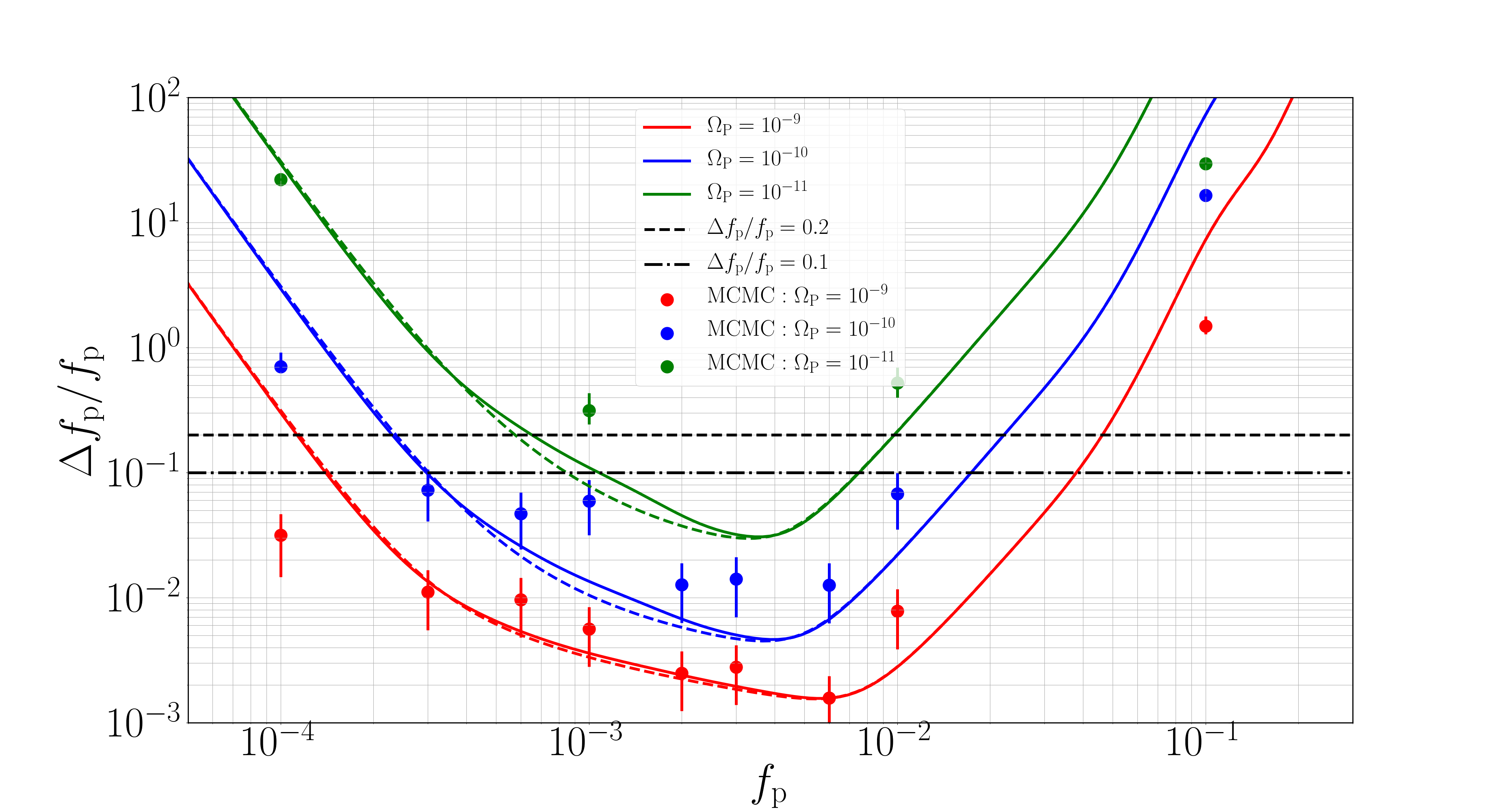}
  \caption{Fisher $\fp$}
  \label{fig:Fisherfp}
\end{subfigure}
\caption{Uncertainty estimates for the peak amplitude $\OmPeak$ and the peak frequency $\fp$, calculated with the Fisher information (continuous lines) and  MCMC simulations (points). The solid line corresponds to a model of LISA instrument noise, phase transition signal, astrophysical background and a  Robson \textit{et al.}~\citep{Robson_2019} DWD foreground model. The dashed lines are identical but instead consider the Lamberts \textit{et al.} model for the DWD foreground \citep{Lamberts:2019nyk}. The relative uncertainties as calculated from the Fisher matrix for  $\OmPeak$ when only the LISA noise and phase transition signal are considered are show in blue in Fig.~\ref{fig:FisherOmega}. In both cases $r_b = 0.4$, $b = 1$ and in Fig.~\ref{fig:FisherOmega} $\fp = 1$ mHz.}
\label{fig:Fisher1}
\end{figure}

\begin{figure}
\begin{subfigure}{.5\textwidth}
  \centering
  \includegraphics[width=\linewidth]{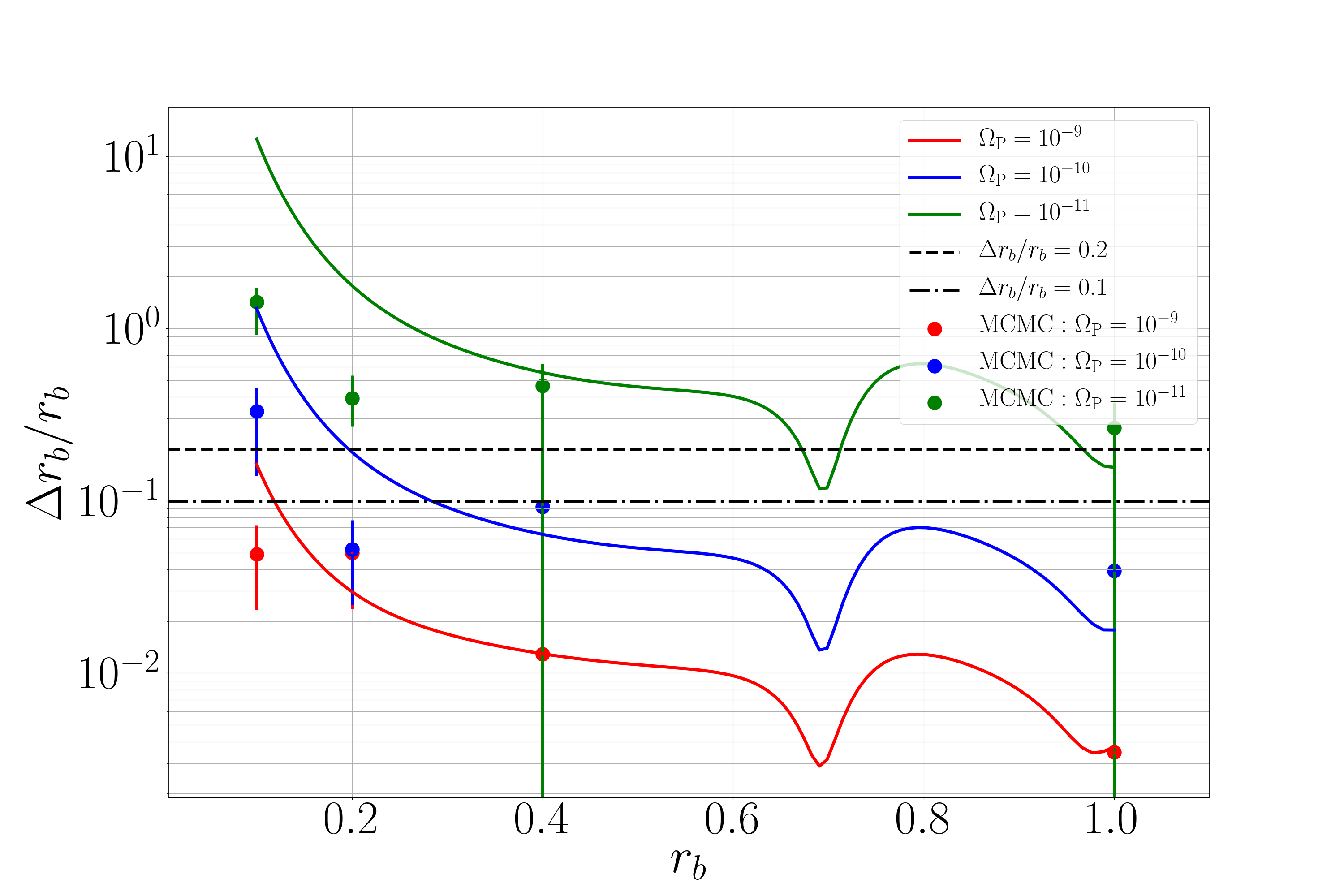}
  \caption{Fisher $r_b$}
  \label{fig:Fisherrb}
\end{subfigure}%
\begin{subfigure}{.5\textwidth}
  \centering
  \includegraphics[width=\linewidth]{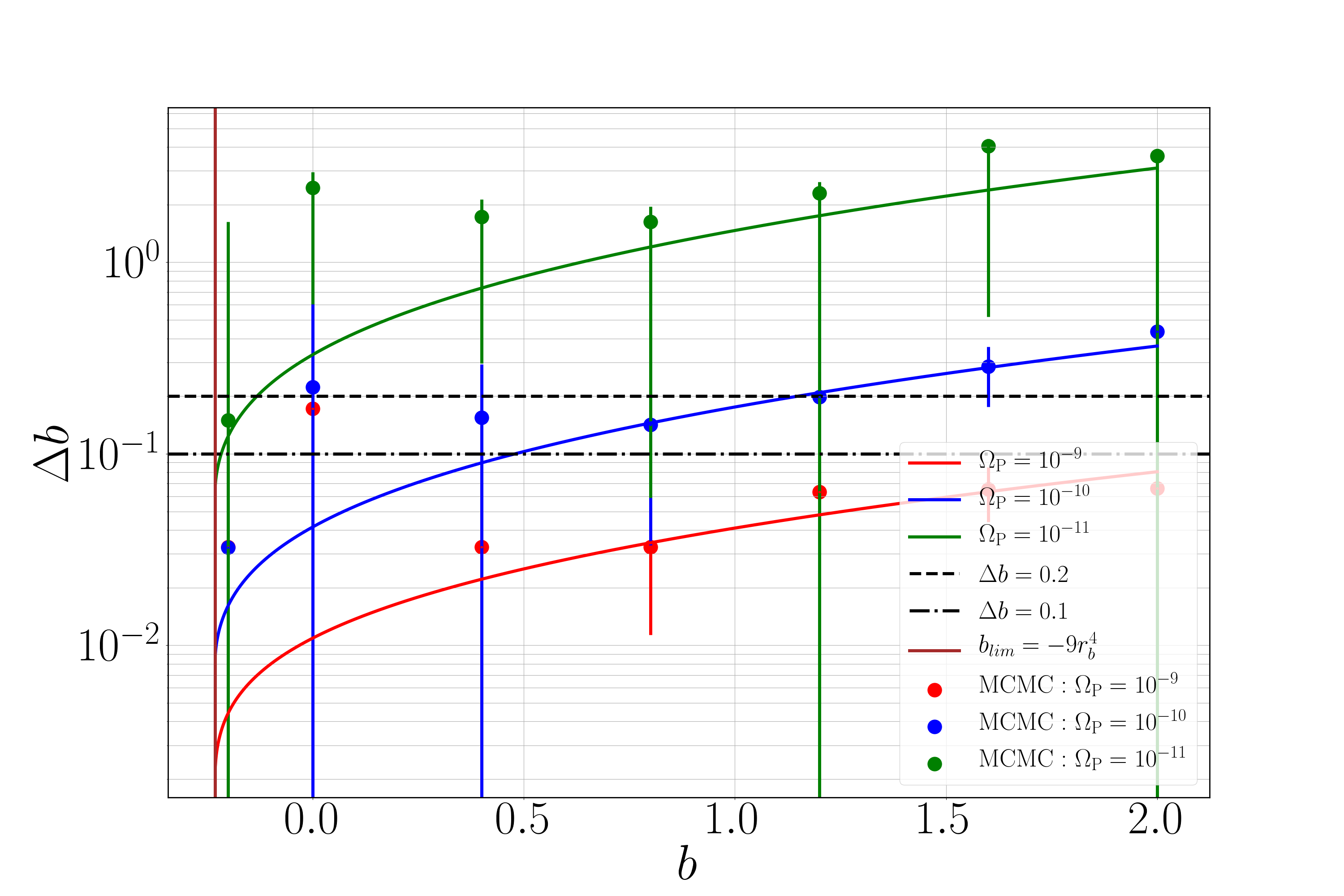}
  \caption{Fisher $b$}
  \label{fig:Fisherb}
\end{subfigure}
\caption{Uncertainty estimates in the break ratio $r_b$ and the intermediate slope $b$, calculated with the Fisher information (continuous lines) and  MCMC simulations (points). The model: LISA instrument noise, first order phase transition signal modelled as a double broken power law, astrophysical background and a Robson \textit{et al.}~\citep{Robson_2019} DWD foreground model. In both figures  $\fp = 1$ mHz, in Fig.~\ref{fig:Fisherrb} $b = 1$ and in Fig.~\ref{fig:Fisherb}  $r_b = 0.4$.}
\label{fig:Fisher2}
\end{figure}

For the MCMC method, we use the same Adaptive-MCMC algorithm as previously presented in~\citep{Boileau:2020rpg,Boileau:2021gbr}. 
We define the uncertainty on a parameter to be the standard deviation of the marginalised posterior distribution. In the following, unless stated otherwise, the total GW model used is described by Eq.~\ref{Eq:OmGWmodel}.
We look at each of the spectral parameters in turn, showing the results for each of the phase transition parameters in Figs.~\ref{fig:FisherOmega}, \ref{fig:Fisherfp}, \ref{fig:Fisherrb} and \ref{fig:Fisherb}. 
The relative uncertainties  calculated from the MCMC results are shown as dots with a 1-$\sigma$ error bar, 
and those from the Fisher information are denoted by continuous lines. 

The Fisher information matrix is much faster to evaluate than an MCMC, and allows us to explore LISA's sensitivity to a wide range of parameter space associated with a first order phase transition; to explore the same parameter space with MCMC methods would take significantly longer.  For this work we aimed to investigate the similarities (and differences) between the Fisher information matrix and MCMC results, which gives insight into how to interpret the Fisher information matrix results for the broader parameter space explored in ~\cite{Gowling:2021gcy}.

In Fig.~\ref{fig:FisherOmega}, the relative uncertainties in the peak amplitude $\OmPeak$ are shown, with the other spectral parameters being $\fp = 1 \times 10^{-3}$ Hz, $r_b = 0.4$ and $b = 1$.  
The agreement between the two ways of estimating the relative uncertainties is very good. 
We also see that for $\OmPeak =  3\times 10^{-11}$, which for this combination of parameters we found to be the threshold for detectability in our DIC analysis (see Fig.~\ref{fig:DICOmega}),  the relative uncertainty of $\Delta \OmPeak/\OmPeak < 0.1$ is reached, consistent with interpreting $\Delta \text{DIC}>5$ as 
a threshold for distinguishing the models.
We see in Fig.~\ref{fig:FisherOmega} that the relative uncertainty decreases as $1/\Omega_p$, before saturating.  As the signal-to-noise ratio (SNR) should be proportional to the peak amplitude $\Omega_p$, this is consistent with the expectation that the relative uncertainty is inversely proportional to the SNR. The saturation occurs when the signal dominates the noise, and there is little further change in the Fisher matrix.

Fig.~\ref{fig:FisherOmega} also shows the impact of the DWD foreground model on the phase  transition measurement. The nature of the DWD foreground spectral density is an open question and as shown in Fig.~22 of \citep{Boileau:2021gbr} the position of the frequency break of the galactic foreground has a large impact on the constraints one is able to place on a flat GW background. 
Due to the computationally intensive nature of MCMC computations we only evaluate the relative uncertainties with the different foreground model using the Fisher matrix. 

The two red lines in Fig.~\ref{fig:FisherOmega} are the two DWD models considered: the dashed line is the analytic fit \citep{Boileau:2021gbr} to the galactic foreground from the Lamberts \textit{et al.} catalogue \citep{Lamberts:2019nyk} (Eq.~\ref{Eq:SGWBDWD}) and the solid line is the galactic confusion noise model from Robson \textit{et al.}~\citep{Robson_2019} (Eq.~\ref{Eq:Cornish_DWD}).  We also show the case where all foregrounds are removed with the blue dashed line; here the signal fluctuations are the sum of the LISA  instrument noise and the fluctuations in the phase transition GW background. 
As in~\citep{Boileau:2021gbr} we see a drop in the limiting performance for low values of $\OmPeak$.  In this case, the drop is larger for the Robson \textit{et al.} model, which can be traced to the model having a higher amplitude at 1 mHz, where the chosen phase transition model power peaks.
The difference in impact of the models decreases with amplitude, as the phase transition signal starts to exceed the power in both foreground models.  In all cases the difference is within the $68\%$ error posterior credible interval, which suggests that the modelling of the DWD foreground is not quite as critical as might be expected.

For the remaining three spectral parameters $\fp$, $r_b$ and $b$,  as well as varying these parameters we have studied the measurement uncertainty coming from the MCMC and Fisher analysis for different values of $\OmPeak$, ($1\times 10^{-9}$, $1\times 10^{-10}$ and $1\times 10^{-11}$) they are shown in green, blue and red respectively in Fig.~\ref{fig:Fisherfp}, Fig.~\ref{fig:Fisherrb} and Fig.~\ref{fig:Fisherb}. It is evident that the uncertainty in all parameters increases when the amplitude is lower. 

In Fig.~\ref{fig:FisherOmega} the solid line displays a total GW model that uses the Robson {\it et al.} galactic foreground model and the dashed line is the analytic model \citep{Boileau:2021gbr} fit on the Galactic foreground from the Lamberts \textit{et al.} catalogue \citep{Lamberts:2019nyk}. As one might expect, 
the peak frequency is less well determined in the louder foreground model; however, the effect is 
not large. It appears that the good overlap between the Fisher study and the Bayesian MCMC analysis disappears when the amplitude $\OmPeak$ decreases. Indeed, in view of the DIC study, when the peak amplitude is below 
$\OmPeak = 3\times 10^{-11}$, we are unable to state with certainty that 
the model including the phase transition signal is a better fit.
It is therefore not surprising that the different methods for estimating the 
parameter uncertainty give different results below this value.

In  Fig.~\ref{fig:Fisherfp} we show the effect of varying the peak frequency, 
for break ratio $r_b = 0.4$ and intermediate slope $b =1$. We see that 
with a peak amplitude $\OmPeak > 1 \times 10^{-10}$, we achieve a relative uncertainty 
$\Delta \fp/\fp < 0.2$ for peak frequency between $\fp = 2 \times 10^{-4}$ Hz and $2\times 10^{-2}$ Hz. We also note an effect of the different galactic models on the measurement of the peak frequency of the phase transition signal.

Finally, in Fig.~\ref{fig:Fisher2}, we display the Fisher and  MCMC results for the two remaining spectral parameters, $r_b$ (the ratio between the breaks in the power laws) and $b$ (the intermediate power law), 
for spectra with peak frequency $\fp = 1$ mHz. 
There is no systematic trend in measurement performance as the parameters are varied, except at low $r_b$, where, for this peak frequency choice,
the lower break frequency moves out of the LISA sensitivity window, and the uncertainty quickly increases.
In Fig.~\ref{fig:Fisherrb} we see a dip in sensitivity at $r_b = 0.7$, this feature is due to the complicated nature of the differential that goes into the Fisher matrix, as opposed to anything special about the spectrum at this combination of spectral parameters. When other parameter combinations are considered  and $r_b$ is varied this dip appears at different $r_b$ values.

For the parameters $r_b$ and $b$, when we compare the parameter estimation results with different galactic foreground models we again see the sensitivity to $r_b$ and $b$ is reduced for the louder galactic foreground model. In Fig.~\ref{fig:Fisherb} the vertical brown line shows the point where the double broken power law, Eq.~\ref{Eq:omgw_dbl_brkn}, becomes ill-defined due to the limitations of the double broken power law discussed in Sec.~\ref{sec:PTfit}.
The relative uncertainty is ill-defined at $b=0$ so here we instead consider $\Delta b$. 

\section{Conclusions}
\label{sec:conclusions}

In this paper we have investigated the ability of LISA to observe a GW background produced by a first order phase transition in the early universe. We have considered the presence of GW foregrounds from DWD binaries in our galaxy, from compact binary mergers throughout the universe, and LISA noise. For a phase transition GW spectrum with break ratio $r_b = 0.4$ and intermediate spectral slope $b =1$,  we show that signals with peak frequency 1 mHz can be detected for $\OmPeak \ge 3\times 10^{-11}$.
Signals with peak amplitude $\OmPeak = 10^{-10}$ achieve the detection threshold $\Delta \text{DIC} > 5$ 
with peak frequency between $\fp = 4 \times 10^{-4}$ to $9 \times10^{-3}$ Hz. For phase transition signals with a larger peak amplitude than $\OmPeak = 10^{-10} $ there would be a broader frequency window of detectability, including phase transition signals with peak frequencies between $ \fp \approx 1 \times 10^{-4}$ to $ \approx 2 \times 10^{-2} $Hz.

We then used Fisher Matrix and MCMC methods to show how well the four parameters associated with the first order phase transition could be estimated, simultaneously with noise and GW foregrounds. For example, with a GW background of peak amplitude $\OmPeak = 10^{-10}$ the parameter estimation accuracies are $\Delta \OmPeak/\OmPeak \approx 10^{-2}$, $\Delta \fp/\fp \approx 10^{-2}$ at $\fp = 3 \times 10^{-3}$,  $\Delta \rb/\rb \approx 0.1$ at $\rb = 0.2$, and $\Delta b/b \approx 0.1$ at $b = 1$. The Fisher Matrix and MCMC methods give similar results for $\OmPeak > 3\times 10^{-11}$, where the signal becomes detectable. 

We have modelled the GW background from a first order phase transition as a double broken power law, which is a good fit to the GW power spectrum calculated from the thermodynamic parameters for the majority of the thermodynamic parameter space. However, subtleties in the characteristics of the GW power spectra from thermodynamic parameters are not encapsulated in the double broken power law, for example the double broken power law struggles to describe the spectra for wall speeds around the speed of sound, see Fig.~11 in \cite{Gowling:2021gcy}. Another challenge for the parameter estimation of first order phase transitions at LISA is, as discussed in Sec.~\ref{sec:PTfit}, the relationship between the spectral and thermodynamic parameters is complicated. See \cite{Gowling:2022pzb} for a discussion on the connection between the spectral and thermodynamic parameters and how to reconstruct thermodynamic parameters from MCMC samples on the spectral parameters (like those performed here). These differences and challenges mean that as we improve our understanding of phase transition physics  and develop better spectral fits, the findings presented here will evolve.  

We have used a basic model for the LISA noise based on only two parameters~\cite{LISA_SR_doc,Smith:2019wny}, and it will be important to incorporate more sophisticated noise models in order to better understand the prospects for cosmological GW background detection and parameter estimation. 

We have been conservative in not using annual modulation to improve the estimation of the DWD foreground parameters. In addition, other GW wave signals will be present in the data, such as identifiable galactic binaries, massive black hole binaries, and extreme mass ratio inspirals~\cite{Amaro-Seoane:2022rxf}. 
Searches may also need to allow for the presence of other cosmological GW backgrounds in the LISA data~\cite{Auclair:2022lcg}, for example from inflation~\cite{Caprini:2018mtu} or cosmic strings~\cite{Auclair:2019wcv,Boileau:2021gbr}. 
More advanced parameter estimation methods will need to be developed, 
and realistic early-universe signal models will need to be included 
in global fits for the LISA GW signals \cite{Babak_2017}.

\section*{Acknowledgements}
G. B. and N. C. thank the Centre national d'\'{e}tudes spatiales for support for this research.
R. M. acknowledges support by the Marsden Fund Council from Government funding, administered by the Royal Society of New Zealand and DFG Grant No. KI 1443/3-. C.G. is supported by a STFC Studentship. MH (ORCID ID 0000-0002-9307-437X) acknowledges support from the Academy of Finland (grant number 333609).

\appendix
\section{Parameter correlations}
\label{app:MCMC_plot}
\begin{figure}
	\centering
	\includegraphics[width=\linewidth]{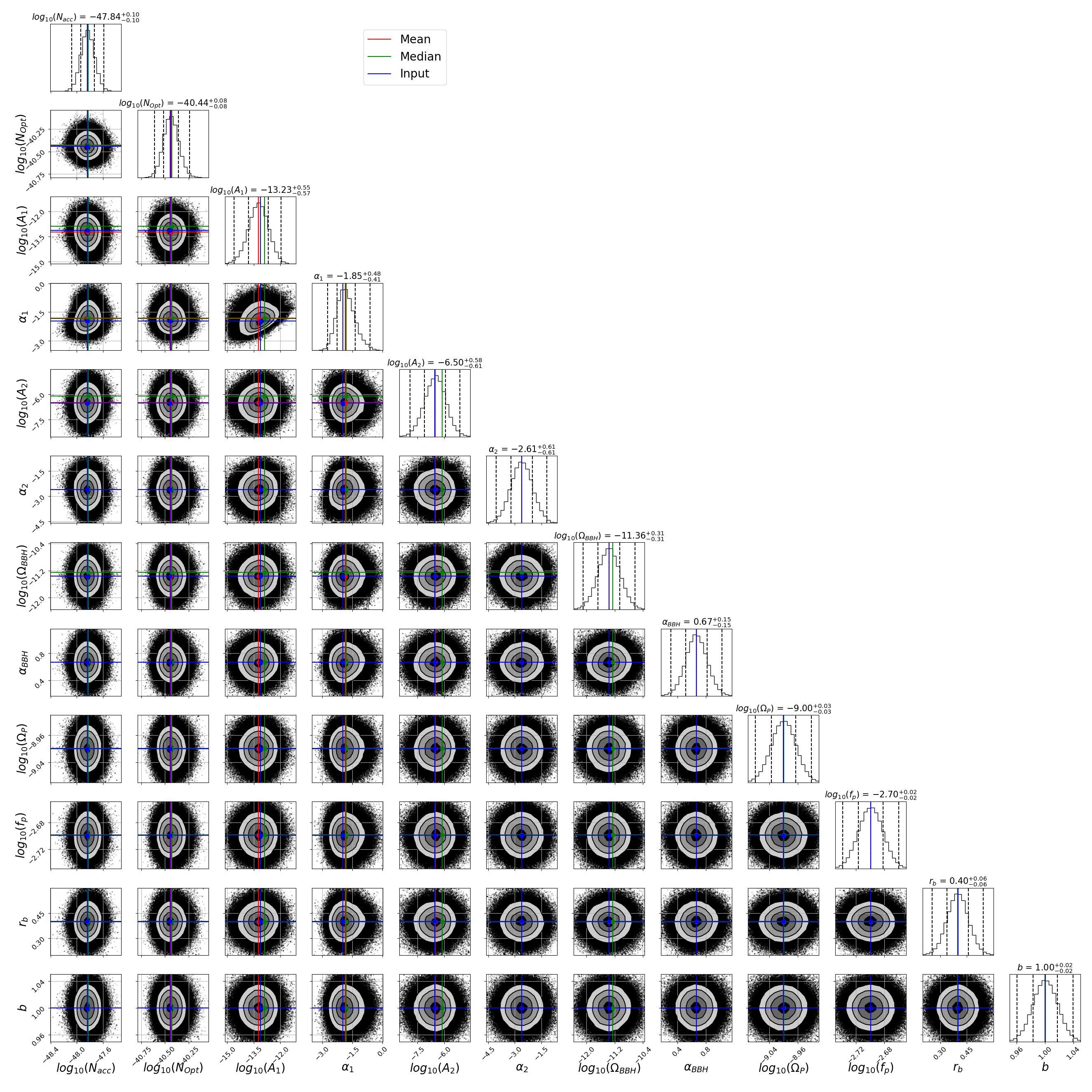}
	\caption{Corner plot for an example adaptive MCMC with an injected phase transition signal characterised by $(\log_{10}(\OmPeak), \log_{10}(f_p), rb, b) = (-9, -2,7, 0.4, 1)$ and a data model described by Eq.~\ref{Eq:OmGWmodel}. The vertical dashed lines on the posterior distribution represent from left to right the quantiles $[16\%,\ 50\%,\ 84\% ]$. The red, green and blues lines are respectively the mean, the median of the posterior distribution and the input parameter values on the simulation.}
	\label{fig:cornerplot}
\end{figure}

When evaluating the Fisher matrix we assumed that parameters from different sources are independent. In  Figure~\ref{fig:cornerplot} we present a corner plot for the MCMC results presented in Section~\ref{sec:Results}. Parameters from different sources can be grouped into different parameter blocks: $\Gamma_{LISA}$, $\Gamma_{ECB}$, $\Gamma_{DWD}$ and $\Gamma_{PT}$. The parameters within a particular block can exhibit small correlations, but parameters from different blocks are independent.

\FloatBarrier
\bibliography{GW_EW_data_analysis}

\providecommand{\href}[2]{#2}\begingroup\raggedright\begin{thebibliography}{10}

\bibitem{Audley:2017drz}
P.~Amaro-Seoane et~al., \emph{{Laser Interferometer Space Antenna}},
  \href{https://arxiv.org/abs/1702.00786}{{\ttfamily 1702.00786}}.

\bibitem{Amaro-Seoane:2022rxf}
P.~Amaro-Seoane et~al., \emph{{Astrophysics with the Laser Interferometer Space
  Antenna}},  \href{https://arxiv.org/abs/2203.06016}{{\ttfamily 2203.06016}}.

\bibitem{Christensen:2018iqi}
N.~Christensen, \emph{{Stochastic Gravitational Wave Backgrounds}},
  \href{https://doi.org/10.1088/1361-6633/aae6b5}{\emph{Rept. Prog. Phys.}
  {\bfseries 82} (2019) 016903}
  [\href{https://arxiv.org/abs/1811.08797}{{\ttfamily 1811.08797}}].

\bibitem{Auclair:2022lcg}
P.~Auclair et~al., \emph{{Cosmology with the Laser Interferometer Space
  Antenna}},  \href{https://arxiv.org/abs/2204.05434}{{\ttfamily 2204.05434}}.

\bibitem{Mazumdar:2018dfl}
A.~Mazumdar and G.~White, \emph{{Review of cosmic phase transitions: their
  significance and experimental signatures}},
  \href{https://doi.org/10.1088/1361-6633/ab1f55}{\emph{Rept. Prog. Phys.}
  {\bfseries 82} (2019) 076901}
  [\href{https://arxiv.org/abs/1811.01948}{{\ttfamily 1811.01948}}].

\bibitem{Hindmarsh:2020hop}
M.B.~Hindmarsh, M.~L\"uben, J.~Lumma and M.~Pauly, \emph{{Phase transitions in
  the early universe}},  \href{https://arxiv.org/abs/2008.09136}{{\ttfamily
  2008.09136}}.

\bibitem{Lamberts:2019nyk}
A.~Lamberts, S.~Blunt, T.B.~Littenberg, S.~Garrison-Kimmel, T.~Kupfer and
  R.E.~Sanderson, \emph{{Predicting the LISA white dwarf binary population in
  the Milky Way with cosmological simulations}},
  \href{https://doi.org/10.1093/mnras/stz2834}{\emph{Mon. Not. Roy. Astron.
  Soc.} {\bfseries 490} (2019) 5888}
  [\href{https://arxiv.org/abs/1907.00014}{{\ttfamily 1907.00014}}].

\bibitem{Adams:2013qma}
M.R.~Adams and N.J.~Cornish, \emph{{Detecting a Stochastic Gravitational Wave
  Background in the presence of a Galactic Foreground and Instrument Noise}},
  \href{https://doi.org/10.1103/PhysRevD.89.022001}{\emph{Phys. Rev. D}
  {\bfseries 89} (2014) 022001}
  [\href{https://arxiv.org/abs/1307.4116}{{\ttfamily 1307.4116}}].

\bibitem{Boileau:2021sni}
G.~Boileau, A.~Lamberts, N.~Christensen, N.J.~Cornish and R.~Meyer,
  \emph{{Spectral separation of the stochastic gravitational-wave background
  for LISA in the context of a modulated Galactic foreground}},
  \href{https://doi.org/10.1093/mnras/stab2575}{\emph{Mon. Not. Roy. Astron.
  Soc.} {\bfseries 508} (2021) 803}
  [\href{https://arxiv.org/abs/2105.04283}{{\ttfamily 2105.04283}}].

\bibitem{KAGRA:2021kbb}
R.~Abbott et~al., \emph{{Upper limits on the isotropic gravitational-wave
  background from Advanced LIGO and Advanced Virgo\textquoteright{}s third
  observing run}},
  \href{https://doi.org/10.1103/PhysRevD.104.022004}{\emph{Phys. Rev. D}
  {\bfseries 104} (2021) 022004}
  [\href{https://arxiv.org/abs/2101.12130}{{\ttfamily 2101.12130}}].

\bibitem{2019ApJ...871...97C}
Z.-C.~{Chen}, F.~{Huang} and Q.-G.~{Huang}, \emph{{Stochastic
  Gravitational-wave Background from Binary Black Holes and Binary Neutron
  Stars and Implications for LISA}},
  \href{https://doi.org/10.3847/1538-4357/aaf581}{\emph{Astrophys. J.}
  {\bfseries 871} (2019) 97}
  [\href{https://arxiv.org/abs/1809.10360}{{\ttfamily 1809.10360}}].

\bibitem{PhysRevD.103.043002}
C.~P\'erigois, C.~Belczynski, T.~Bulik and T.~Regimbau, \emph{Startrack
  predictions of the stochastic gravitational-wave background from compact
  binary mergers},
  \href{https://doi.org/10.1103/PhysRevD.103.043002}{\emph{Phys. Rev. D}
  {\bfseries 103} (2021) 043002}.

\bibitem{Kajantie:1996mn}
K.~Kajantie, M.~Laine, K.~Rummukainen and M.E.~Shaposhnikov, \emph{{Is there a
  hot electroweak phase transition at m(H) larger or equal to m(W)?}},
  \href{https://doi.org/10.1103/PhysRevLett.77.2887}{\emph{Phys. Rev. Lett.}
  {\bfseries 77} (1996) 2887}
  [\href{https://arxiv.org/abs/hep-ph/9605288}{{\ttfamily hep-ph/9605288}}].

\bibitem{Kajantie:1996qd}
K.~Kajantie, M.~Laine, K.~Rummukainen and M.E.~Shaposhnikov, \emph{{A
  Nonperturbative analysis of the finite T phase transition in SU(2) x U(1)
  electroweak theory}},
  \href{https://doi.org/10.1016/S0550-3213(97)00164-8}{\emph{Nucl. Phys. B}
  {\bfseries 493} (1997) 413}
  [\href{https://arxiv.org/abs/hep-lat/9612006}{{\ttfamily hep-lat/9612006}}].

\bibitem{Caprini:2019egz}
C.~Caprini et~al., \emph{{Detecting gravitational waves from cosmological phase
  transitions with LISA: an update}},
  \href{https://doi.org/10.1088/1475-7516/2020/03/024}{\emph{JCAP} {\bfseries
  03} (2020) 024} [\href{https://arxiv.org/abs/1910.13125}{{\ttfamily
  1910.13125}}].

\bibitem{Hindmarsh:2013xza}
M.~Hindmarsh, S.J.~Huber, K.~Rummukainen and D.J.~Weir, \emph{{Gravitational
  waves from the sound of a first order phase transition}},
  \href{https://doi.org/10.1103/PhysRevLett.112.041301}{\emph{Phys. Rev. Lett.}
  {\bfseries 112} (2014) 041301}
  [\href{https://arxiv.org/abs/1304.2433}{{\ttfamily 1304.2433}}].

\bibitem{Hindmarsh:2015qta}
M.~Hindmarsh, S.J.~Huber, K.~Rummukainen and D.J.~Weir, \emph{{Numerical
  simulations of acoustically generated gravitational waves at a first order
  phase transition}},
  \href{https://doi.org/10.1103/PhysRevD.92.123009}{\emph{Phys. Rev. D}
  {\bfseries 92} (2015) 123009}
  [\href{https://arxiv.org/abs/1504.03291}{{\ttfamily 1504.03291}}].

\bibitem{Hindmarsh:2017gnf}
M.~Hindmarsh, S.J.~Huber, K.~Rummukainen and D.J.~Weir, \emph{{Shape of the
  acoustic gravitational wave power spectrum from a first order phase
  transition}}, \href{https://doi.org/10.1103/PhysRevD.96.103520}{\emph{Phys.
  Rev. D} {\bfseries 96} (2017) 103520}
  [\href{https://arxiv.org/abs/1704.05871}{{\ttfamily 1704.05871}}].

\bibitem{Cutting:2018tjt}
D.~Cutting, M.~Hindmarsh and D.J.~Weir, \emph{{Gravitational waves from vacuum
  first-order phase transitions: from the envelope to the lattice}},
  \href{https://doi.org/10.1103/PhysRevD.97.123513}{\emph{Phys. Rev. D}
  {\bfseries 97} (2018) 123513}
  [\href{https://arxiv.org/abs/1802.05712}{{\ttfamily 1802.05712}}].

\bibitem{Cutting:2020nla}
D.~Cutting, E.G.~Escartin, M.~Hindmarsh and D.J.~Weir, \emph{{Gravitational
  waves from vacuum first order phase transitions II: from thin to thick
  walls}},  \href{https://arxiv.org/abs/2005.13537}{{\ttfamily 2005.13537}}.

\bibitem{Lewicki:2020azd}
M.~Lewicki and V.~Vaskonen, \emph{{Gravitational waves from colliding vacuum
  bubbles in gauge theories}},
  \href{https://doi.org/10.1140/epjc/s10052-021-09232-3}{\emph{Eur. Phys. J. C}
  {\bfseries 81} (2021) 437}
  [\href{https://arxiv.org/abs/2012.07826}{{\ttfamily 2012.07826}}].

\bibitem{Lewicki:2020jiv}
M.~Lewicki and V.~Vaskonen, \emph{{Gravitational wave spectra from strongly
  supercooled phase transitions}},
  \href{https://doi.org/10.1140/epjc/s10052-020-08589-1}{\emph{Eur. Phys. J. C}
  {\bfseries 80} (2020) 1003}
  [\href{https://arxiv.org/abs/2007.04967}{{\ttfamily 2007.04967}}].

\bibitem{Lee:2021nwg}
B.-H.~Lee, W.~Lee, D.-h.~Yeom and L.~Yin, \emph{{Gravitational waves from the
  vacuum decay with LISA *}},
  \href{https://doi.org/10.1088/1674-1137/ac5d2a}{\emph{Chin. Phys. C}
  {\bfseries 46} (2022) 075101}
  [\href{https://arxiv.org/abs/2106.07430}{{\ttfamily 2106.07430}}].

\bibitem{Lewicki:2022pdb}
M.~Lewicki and V.~Vaskonen, \emph{{Gravitational waves from bubble collisions
  and fluid motion in strongly supercooled phase transitions}},
  \href{https://arxiv.org/abs/2208.11697}{{\ttfamily 2208.11697}}.

\bibitem{Ellis:2019oqb}
J.~Ellis, M.~Lewicki, J.M.~No and V.~Vaskonen, \emph{{Gravitational wave energy
  budget in strongly supercooled phase transitions}},
  \href{https://doi.org/10.1088/1475-7516/2019/06/024}{\emph{JCAP} {\bfseries
  06} (2019) 024} [\href{https://arxiv.org/abs/1903.09642}{{\ttfamily
  1903.09642}}].

\bibitem{Ellis:2020nnr}
J.~Ellis, M.~Lewicki and V.~Vaskonen, \emph{{Updated predictions for
  gravitational waves produced in a strongly supercooled phase transition}},
  \href{https://doi.org/10.1088/1475-7516/2020/11/020}{\emph{JCAP} {\bfseries
  11} (2020) 020} [\href{https://arxiv.org/abs/2007.15586}{{\ttfamily
  2007.15586}}].

\bibitem{Hindmarsh:2016lnk}
M.~Hindmarsh, \emph{{Sound shell model for acoustic gravitational wave
  production at a first-order phase transition in the early Universe}},
  \href{https://doi.org/10.1103/PhysRevLett.120.071301}{\emph{Phys. Rev. Lett.}
  {\bfseries 120} (2018) 071301}
  [\href{https://arxiv.org/abs/1608.04735}{{\ttfamily 1608.04735}}].

\bibitem{Hindmarsh:2019phv}
M.~Hindmarsh and M.~Hijazi, \emph{{Gravitational waves from first order
  cosmological phase transitions in the Sound Shell Model}},
  \href{https://doi.org/10.1088/1475-7516/2019/12/062}{\emph{JCAP} {\bfseries
  12} (2019) 062} [\href{https://arxiv.org/abs/1909.10040}{{\ttfamily
  1909.10040}}].

\bibitem{Cutting:2019zws}
D.~Cutting, M.~Hindmarsh and D.J.~Weir, \emph{{Vorticity, kinetic energy, and
  suppressed gravitational wave production in strong first order phase
  transitions}},
  \href{https://doi.org/10.1103/PhysRevLett.125.021302}{\emph{Phys. Rev. Lett.}
  {\bfseries 125} (2020) 021302}
  [\href{https://arxiv.org/abs/1906.00480}{{\ttfamily 1906.00480}}].

\bibitem{Dahl:2021wyk}
J.~Dahl, M.~Hindmarsh, K.~Rummukainen and D.J.~Weir, \emph{{Decay of acoustic
  turbulence in two dimensions and implications for cosmological gravitational
  waves}}, \href{https://doi.org/10.1103/PhysRevD.106.063511}{\emph{Phys. Rev.
  D} {\bfseries 106} (2022) 063511}
  [\href{https://arxiv.org/abs/2112.12013}{{\ttfamily 2112.12013}}].

\bibitem{Auclair:2022jod}
P.~Auclair, C.~Caprini, D.~Cutting, M.~Hindmarsh, K.~Rummukainen, D.A.~Steer
  et~al., \emph{{Generation of gravitational waves from freely decaying
  turbulence}},
  \href{https://doi.org/10.1088/1475-7516/2022/09/029}{\emph{JCAP} {\bfseries
  09} (2022) 029} [\href{https://arxiv.org/abs/2205.02588}{{\ttfamily
  2205.02588}}].

\bibitem{Giese:2020rtr}
F.~Giese, T.~Konstandin and J.~van~de Vis, \emph{{Model-independent energy
  budget of cosmological first-order phase transitions\textemdash{}A sound
  argument to go beyond the bag model}},
  \href{https://doi.org/10.1088/1475-7516/2020/07/057}{\emph{JCAP} {\bfseries
  07} (2020) 057} [\href{https://arxiv.org/abs/2004.06995}{{\ttfamily
  2004.06995}}].

\bibitem{Giese:2020znk}
F.~Giese, T.~Konstandin, K.~Schmitz and J.~Van De~Vis, \emph{{Model-independent
  energy budget for LISA}},
  \href{https://doi.org/10.1088/1475-7516/2021/01/072}{\emph{JCAP} {\bfseries
  01} (2021) 072} [\href{https://arxiv.org/abs/2010.09744}{{\ttfamily
  2010.09744}}].

\bibitem{Gowling:2021gcy}
C.~Gowling and M.~Hindmarsh, \emph{{Observational prospects for phase
  transitions at LISA: Fisher matrix analysis}},
  \href{https://arxiv.org/abs/2106.05984}{{\ttfamily 2106.05984}}.

\bibitem{Giese:2021dnw}
F.~Giese, T.~Konstandin and J.~van~de Vis, \emph{{Finding sound shells in LISA
  mock data using likelihood sampling}},
  \href{https://doi.org/10.1088/1475-7516/2021/11/002}{\emph{JCAP} {\bfseries
  11} (2021) 002} [\href{https://arxiv.org/abs/2107.06275}{{\ttfamily
  2107.06275}}].

\bibitem{Spiegelhalter2002}
D.J.~Spiegelhalter, N.G.~Best, B.P.~Carlin and A.~van~der Linde,
  \emph{{Bayesian measures of model complexity and fit}},
  \href{https://doi.org/10.1111/1467-9868.00353}{\emph{J. Roy. Statist. Soc. B}
  {\bfseries 64} (2002) 583}.

\bibitem{Spiegelhalter2014}
D.J.~Spiegelhalter, N.G.~Best, B.P.~Carlin and A.~van~der Linde, \emph{The
  deviance information criterion: 12 years on}, {\emph{Journal of the Royal
  Statistical Society. Series B (Statistical Methodology)} {\bfseries 76}
  (2014) 485}.

\bibitem{MeyerDICreview2016}
R.~Meyer, \emph{Deviance information criterion (dic)},  in \emph{Wiley
  StatsRef: Statistics Reference Online}, pp.~1--6, Wiley \& Sons (2016),
  \href{https://doi.org/https://doi.org/10.1002/9781118445112.stat07878}{DOI}.

\bibitem{Tinto:2001ii}
M.~Tinto, J.W.~Armstrong and F.B.~Estabrook, \emph{{Discriminating a
  gravitational wave background from instrumental noise in the LISA detector}},
  \href{https://doi.org/10.1103/PhysRevD.63.021101}{\emph{Phys. Rev. D}
  {\bfseries 63} (2001) 021101}.

\bibitem{Tinto:2002de}
M.~Tinto, F.B.~Estabrook and J.W.~Armstrong, \emph{{Time delay interferometry
  for LISA}}, \href{https://doi.org/10.1103/PhysRevD.65.082003}{\emph{Phys.
  Rev. D} {\bfseries 65} (2002) 082003}.

\bibitem{Vallisneri_2012}
M.~Vallisneri and C.R.~Galley, \emph{Non-sky-averaged sensitivity curves for
  space-based gravitational-wave observatories},
  \href{https://doi.org/10.1088/0264-9381/29/12/124015}{\emph{Classical and
  Quantum Gravity} {\bfseries 29} (2012) 124015}.

\bibitem{Flauger_2021}
R.~Flauger, N.~Karnesis, G.~Nardini, M.~Pieroni, A.~Ricciardone and J.~Torrado,
  \emph{Improved reconstruction of a stochastic gravitational wave background
  with {LISA}},
  \href{https://doi.org/10.1088/1475-7516/2021/01/059}{\emph{Journal of
  Cosmology and Astroparticle Physics} {\bfseries 2021} (2021) 059}.

\bibitem{Smith:2019wny}
T.L.~Smith and R.~Caldwell, \emph{{LISA for Cosmologists: Calculating the
  Signal-to-Noise Ratio for Stochastic and Deterministic Sources}},
  \href{https://doi.org/10.1103/PhysRevD.100.104055}{\emph{Phys. Rev. D}
  {\bfseries 100} (2019) 104055}
  [\href{https://arxiv.org/abs/1908.00546}{{\ttfamily 1908.00546}}].

\bibitem{LISA_SR_doc}
\emph{{LISA Science Requirements Document}},  Tech. Rep.
  \href{https://www.cosmos.esa.int/web/lisa/lisa-documents}{ESA-L3-EST-SCI-RS-001}
  (2019).

\bibitem{Baker:2019nia}
J.~Baker et~al., \emph{{The Laser Interferometer Space Antenna: Unveiling the
  Millihertz Gravitational Wave Sky}},
  \href{https://arxiv.org/abs/1907.06482}{{\ttfamily 1907.06482}}.

\bibitem{Nelemans_2001}
G.~Nelemans, L.R.~Yungelson, S.F.~Portegies~Zwart and F.~Verbunt,
  \emph{Population synthesis for double white dwarfs},
  \href{https://doi.org/10.1051/0004-6361:20000147}{\emph{Astronomy \&
  Astrophysics} {\bfseries 365} (2001) 491}.

\bibitem{PhysRevD.76.083006}
N.J.~Cornish and T.B.~Littenberg, \emph{Tests of bayesian model selection
  techniques for gravitational wave astronomy},
  \href{https://doi.org/10.1103/PhysRevD.76.083006}{\emph{Phys. Rev. D}
  {\bfseries 76} (2007) 083006}.

\bibitem{Ruiter_2010}
A.J.~Ruiter, K.~Belczynski, M.~Benacquista, S.L.~Larson and G.~Williams,
  \emph{{The LISA Gravitational Wave Foreground: A Study of Double White
  Dwarfs}},
  \href{https://doi.org/10.1088/0004-637X/717/2/1006}{\emph{Astrophys. J.}
  {\bfseries 717} (2010) 1006}
  [\href{https://arxiv.org/abs/0705.3272}{{\ttfamily 0705.3272}}].

\bibitem{2014PhRvD..89b2001A}
M.R.~{Adams} and N.J.~{Cornish}, \emph{{Detecting a stochastic gravitational
  wave background in the presence of a galactic foreground and instrument
  noise}}, \href{https://doi.org/10.1103/PhysRevD.89.022001}{\emph{\prd}
  {\bfseries 89} (2014) 022001}
  [\href{https://arxiv.org/abs/1307.4116}{{\ttfamily 1307.4116}}].

\bibitem{2017PASA...34...58E}
J.J.~{Eldridge}, E.R.~{Stanway}, L.~{Xiao}, L.A.S.~{McClelland}, G.~{Taylor},
  M.~{Ng} et~al., \emph{{Binary Population and Spectral Synthesis Version 2.1:
  Construction, Observational Verification, and New Results}},
  \href{https://doi.org/10.1017/pasa.2017.51}{\emph{\pasa} {\bfseries 34}
  (2017) e058} [\href{https://arxiv.org/abs/1710.02154}{{\ttfamily
  1710.02154}}].

\bibitem{Hernandez_2020}
M.S.~Hernandez, M.R.~Schreiber, S.G.~Parsons, B.T.~G{\"a}nsicke, F.~Lagos,
  R.~Raddi et~al., \emph{The white dwarf binary pathways survey -- iv. three
  close white dwarf binaries with g-type secondary stars},
  \href{https://doi.org/10.1093/mnras/staa3815}{\emph{Monthly Notices of the
  Royal Astronomical Society} {\bfseries 501} (2020) 1677}.

\bibitem{PhysRevD.64.121501}
C.~Ungarelli and A.~Vecchio, \emph{Studying the anisotropy of the gravitational
  wave stochastic background with lisa},
  \href{https://doi.org/10.1103/PhysRevD.64.121501}{\emph{Phys. Rev. D}
  {\bfseries 64} (2001) 121501}.

\bibitem{Boileau:2021gbr}
G.~Boileau, A.C.~Jenkins, M.~Sakellariadou, R.~Meyer and N.~Christensen,
  \emph{{Ability of LISA to detect a gravitational-wave background of
  cosmological origin: The cosmic string case}},
  \href{https://doi.org/10.1103/PhysRevD.105.023510}{\emph{Phys. Rev. D}
  {\bfseries 105} (2022) 023510}
  [\href{https://arxiv.org/abs/2109.06552}{{\ttfamily 2109.06552}}].

\bibitem{Robson_2019}
T.~Robson, N.J.~Cornish and C.~Liu, \emph{The construction and use of {LISA}
  sensitivity curves},
  \href{https://doi.org/10.1088/1361-6382/ab1101}{\emph{Classical and Quantum
  Gravity} {\bfseries 36} (2019) 105011}.

\bibitem{Karnesis:2021tsh}
N.~Karnesis, S.~Babak, M.~Pieroni, N.~Cornish and T.~Littenberg,
  \emph{{Characterization of the stochastic signal originating from compact
  binary populations as measured by LISA}},
  \href{https://doi.org/10.1103/PhysRevD.104.043019}{\emph{Phys. Rev. D}
  {\bfseries 104} (2021) 043019}
  [\href{https://arxiv.org/abs/2103.14598}{{\ttfamily 2103.14598}}].

\bibitem{Chen:2018rzo}
Z.-C.~Chen, F.~Huang and Q.-G.~Huang, \emph{{Stochastic Gravitational-wave
  Background from Binary Black Holes and Binary Neutron Stars and Implications
  for LISA}}, \href{https://doi.org/10.3847/1538-4357/aaf581}{\emph{Astrophys.
  J.} {\bfseries 871} (2019) 97}
  [\href{https://arxiv.org/abs/1809.10360}{{\ttfamily 1809.10360}}].

\bibitem{Farmer:2003pa}
A.J.~Farmer and E.S.~Phinney, \emph{{The gravitational wave background from
  cosmological compact binaries}},
  \href{https://doi.org/10.1111/j.1365-2966.2003.07176.x}{\emph{Mon. Not. Roy.
  Astron. Soc.} {\bfseries 346} (2003) 1197}
  [\href{https://arxiv.org/abs/astro-ph/0304393}{{\ttfamily
  astro-ph/0304393}}].

\bibitem{LIGOScientific:2019vic}
B.P.~Abbott et~al., \emph{{Search for the isotropic stochastic background using
  data from Advanced LIGO\textquoteright{}s second observing run}},
  \href{https://doi.org/10.1103/PhysRevD.100.061101}{\emph{Phys. Rev. D}
  {\bfseries 100} (2019) 061101}
  [\href{https://arxiv.org/abs/1903.02886}{{\ttfamily 1903.02886}}].

\bibitem{Babak:2021mhe}
S.~Babak, A.~Petiteau and M.~Hewitson, \emph{{LISA Sensitivity and SNR
  Calculations}},  \href{https://arxiv.org/abs/2108.01167}{{\ttfamily
  2108.01167}}.

\bibitem{Cornish:2018dyw}
T.~Robson, N.J.~Cornish and C.~Liu, \emph{{The construction and use of LISA
  sensitivity curves}},
  \href{https://doi.org/10.1088/1361-6382/ab1101}{\emph{Class. Quant. Grav.}
  {\bfseries 36} (2019) 105011}
  [\href{https://arxiv.org/abs/1803.01944}{{\ttfamily 1803.01944}}].

\bibitem{PhysRevD.92.063002}
D.~Meacher, M.~Coughlin, S.~Morris, T.~Regimbau, N.~Christensen, S.~Kandhasamy
  et~al., \emph{Mock data and science challenge for detecting an astrophysical
  stochastic gravitational-wave background with advanced ligo and advanced
  virgo}, \href{https://doi.org/10.1103/PhysRevD.92.063002}{\emph{Phys. Rev. D}
  {\bfseries 92} (2015) 063002}.

\bibitem{https://doi.org/10.48550/arxiv.2205.00416}
K.~Janssens, G.~Boileau, M.-A.~Bizouard, N.~Christensen, T.~Regimbau and N.~van
  Remortel, \emph{Formalism for power spectral density estimation for
  non-identical and correlated noise using the null channel in einstein
  telescope},  2022.
\newblock 10.48550/ARXIV.2205.00416.

\bibitem{doi:10.1198/jcgs.2009.06134}
G.O.~Roberts and J.S.~Rosenthal, \emph{Examples of adaptive mcmc},
  \href{https://doi.org/10.1198/jcgs.2009.06134}{\emph{Journal of Computational
  and Graphical Statistics} {\bfseries 18} (2009) 349}
  [\href{https://arxiv.org/abs/https://doi.org/10.1198/jcgs.2009.06134}{{\ttfamily
  https://doi.org/10.1198/jcgs.2009.06134}}].

\bibitem{LindleyD.V.1977Apif}
D.~Lindley, \emph{A problem in forensic science}, {\emph{Biometrika} {\bfseries
  64} (1977) 207}.

\bibitem{ShaferGlenn1982LP}
G.~Shafer, \emph{Lindley's paradox}, {\emph{Journal of the American Statistical
  Association} {\bfseries 77} (1982) 325}.

\bibitem{MAURI2016570}
C.~Mauri and M.G.~Paris, \emph{The lindley paradox in optical interferometry},
  \href{https://doi.org/https://doi.org/10.1016/j.physleta.2015.11.040}{\emph{Physics
  Letters A} {\bfseries 380} (2016) 570}.

\bibitem{LunnDavid2013TBb}
D.~Lunn, \emph{{The BUGS Book: A Practical Introduction to Bayesian Analysis}},
  Texts in statistical science (2013).

\bibitem{Gowling:2022pzb}
C.~Gowling, M.~Hindmarsh, D.C.~Hooper and J.~Torrado, \emph{{Reconstructing
  physical parameters from template gravitational wave spectra at LISA: first
  order phase transitions}},
  \href{https://arxiv.org/abs/2209.13551}{{\ttfamily 2209.13551}}.

\bibitem{Boileau:2020rpg}
G.~Boileau, N.~Christensen, R.~Meyer and N.J.~Cornish, \emph{{Spectral
  separation of the stochastic gravitational-wave background for LISA:
  Observing both cosmological and astrophysical backgrounds}},
  \href{https://doi.org/10.1103/PhysRevD.103.103529}{\emph{Phys. Rev. D}
  {\bfseries 103} (2021) 103529}
  [\href{https://arxiv.org/abs/2011.05055}{{\ttfamily 2011.05055}}].

\bibitem{Caprini:2018mtu}
C.~Caprini and D.G.~Figueroa, \emph{{Cosmological Backgrounds of Gravitational
  Waves}}, \href{https://doi.org/10.1088/1361-6382/aac608}{\emph{Class. Quant.
  Grav.} {\bfseries 35} (2018) 163001}
  [\href{https://arxiv.org/abs/1801.04268}{{\ttfamily 1801.04268}}].

\bibitem{Auclair:2019wcv}
P.~Auclair et~al., \emph{{Probing the gravitational wave background from cosmic
  strings with LISA}},
  \href{https://doi.org/10.1088/1475-7516/2020/04/034}{\emph{JCAP} {\bfseries
  04} (2020) 034} [\href{https://arxiv.org/abs/1909.00819}{{\ttfamily
  1909.00819}}].

\bibitem{Babak_2017}
S.~Babak, \emph{{\textquotedblleft}enchilada{\textquotedblright} is back on the
  menu}, \href{https://doi.org/10.1088/1742-6596/840/1/012026}{\emph{Journal of
  Physics: Conference Series} 012026}.

\end{thebibliography}\endgroup

\end{document}